\documentclass[onecolumn]{emulateapj}

\newcommand{\wcap}{0.476\textwidth}
\newcommand{\gray}{$\gamma$-ray}
\newcommand{\grays}{$\gamma$-rays}

\newcommand{\hi}{H {\sc i}}

\newcommand{\Xco}{$X_{\rm CO}$}

\newcommand{\icrc}{Int.\ Cosmic Ray Conf.\ }
\newcommand{\adv}{Adv.\ Space Res.}

\newcommand{\pr}{Phys.\ Rev.}
\newcommand{\plb}{Phys.\ Lett.\ B}
\newcommand{\pubjournal}[5]{#4, #1, #2, #3}
\newcommand{\pubjournala}[5]{#4, #1, #3}
\newcommand{\pubjournalb}[5]{#4, #5 #1 #3}

\shorttitle{Dissipation of MHD waves}
\shortauthors{Ptuskin et al.}

\begin{document}

\title{Dissipation of Magnetohydrodynamic Waves on Energetic Particles:
Impact on Interstellar Turbulence and Cosmic Ray Transport}

\author{Vladimir S. Ptuskin\altaffilmark{1}}
\affil{Institute of Terrestrial Magnetism,
Ionosphere and Radio Wave Propagation (IZMIRAN), Troitsk, Moscow region
142190, Russia
\email{vptuskin@izmiran.rssi.ru}}
\altaffiltext{1}{Also Institute for Physical Science and Technology,
University of Maryland, College Park, MD 20742}

\author{Igor V. Moskalenko\altaffilmark{2}}
\affil{
   Hansen Experimental Physics Laboratory, 
   Stanford University, Stanford, CA 94305
\email{imos@stanford.edu}}
\altaffiltext{2}{Also Kavli Institute for Particle Astrophysics and Cosmology,
Stanford University, Stanford, CA 94309}

\author{Frank C. Jones}
\affil{Exploration of the Universe Division, 
NASA/Goddard Space Flight Center, Code 661,
Greenbelt, MD 20771
\email{Frank.C.Jones@gsfc.nasa.gov}}

\author{Andrew W. Strong}
\affil{Max-Planck-Institut f\"ur extraterrestrische Physik,
Postfach 1312, D-85741 Garching, Germany
\email{aws@mpe.mpg.de}}

\and

\author{Vladimir N. Zirakashvili\altaffilmark{3}}
\affil{Institute of Terrestrial Magnetism,
Ionosphere and Radio Wave Propagation (IZMIRAN), Troitsk, Moscow region
142190, Russia \email{zirak@mpimail.mpi-hd.mpg.de}}
\altaffiltext{3}{Also Max-Planck-Institut fur Kernphysik, D-69029,
Heidelberg, Postfach 103980, Germany}

\begin{abstract}
The physical processes involved in diffusion of Galactic cosmic rays in 
the interstellar medium are addressed.
We study the possibility  that the nonlinear MHD cascade sets the
power-law spectrum of turbulence which scatters charged energetic particles.
We find that the dissipation of waves  due to the resonant interaction
with cosmic ray particles may terminate the Kraichnan-type cascade below wavelengths 
$10^{13}$ cm.
The effect of this wave dissipation has been incorporated in the GALPROP numerical 
propagation code in order to asses the impact on measurable astrophysical data.
The energy-dependence of the cosmic-ray diffusion coefficient 
found in the resulting self-consistent model may explain the peaks in the secondary
to primary nuclei ratios observed at about 1 GeV/nucleon.

\end{abstract}


\section{Introduction}

The Galactic cosmic rays -- the gas of relativistic charged particles with high
energy density -- cannot always be  treated as test particles moving in given
magnetic fields. In particular, the stochastic acceleration of cosmic rays by
MHD waves is accompanied by the damping of the waves, since the wave energy is
dissipated. The rate of wave damping on cosmic rays through the cyclotron
resonance interaction was first estimated by \citet{Tid66}. If we exclude cold HI
regions (where the waves are damped by collisions of ions with neutral atoms)
and also perhaps regions of the interstellar gas with very high temperature
$T\sim10^{6}$ K and weak magnetic field (where Landau damping on thermal
particles is high), then one  finds that this mechanism of dissipation could
dominate over other known mechanisms in the interval of wavelengths $10^{11}$
cm to $10^{14}$ cm. The non-resonant interaction of diffusing cosmic rays with
magnetosonic waves \citep{Ptu81} is important only at large wavelengths and is
not important for the present investigation. 
We also do not consider the large body of instabilities in cosmic rays that may
arise because of the non-equilibrium distribution of charged energetic particles
caused by possible strong anisotropy or large gradients and which may amplify waves
in the background plasma -- see e.g.\ \citet{Ber90} and
\citet{Die01} for a review of such processes.  Cyclotron wave damping on cosmic rays
changes the wave spectrum in the interstellar medium, which in turn affects the
particle transport since the cosmic ray diffusion coefficient is determined by
the level of turbulence which scatters charged particles. Thus in principle
the study of cosmic ray diffusion requires a self-consistent approach. We
shall see below that the effect of cosmic rays on interstellar turbulence should
be taken into account at energies below a few GeV/nucleon, and this  may
result in a considerable increase in the  particle diffusion coefficient. This
picture will be also verified by its  consistency with  observations of
interstellar turbulence. Consistency of this picture with observations of
interstellar turbulence is also discussed

It is remarkable that the interpretation of cosmic ray data on secondary
nuclei may require this effect. The secondary nuclei are produced in cosmic rays in
the course of diffusion and nuclear interactions of primary nuclei with
interstellar gas. The $^{2}$H, $^{3}$He, Li, Be, B and many 
other isotopes and elements are almost pure secondaries. Their abundance in
cosmic ray sources is negligible and they result from the fragmentation of
 heavier nuclei. Cosmic ray antiprotons and the major fraction of
positrons are among the secondary species as well. The ratio of secondary to
primary nuclei such as \textrm{B/C} has a peak at about $1$ GeV/n and decreases
 both with increasing and decreasing energy, e.g. \citet{Eng90}.
The high energy behavior is naturally explained by  particle scattering on
 interstellar MHD turbulence with a power-law spectrum in wavelength, but
the required sharp increase of the diffusion coefficient at small energies has
usually been considered as improbable since it needs a drastic ``physically
unjustified'' bending down of the wave spectrum at wave numbers $k>1/3\times
10^{12}$ cm$^{-1}$, see e.g.\ \citet{Jon01},
\citet{M02}.  Alternative explanations of the peaks in secondary to primary
ratios without invoking peculiarities in the
cosmic ray diffusion coefficient have been
suggested. The two most popular of these involve diffusive-convective
particle transport in the hypothetical Galactic wind \citep{Jon79,Ptu97}
and the stochastic reacceleration of cosmic rays by interstellar
turbulence \citep{Sim86,Seo94}. 
These processes are included in the numerical computations of cosmic ray
transport in the Galaxy in the frameworks of the GALPROP code \citep{Str98,M02}. The
situation with the interpretation of the energy dependence of secondaries in
cosmic rays is still uncertain. The peaks in the secondary to primary nuclei
ratios calculated in  diffusion-convection models seem to be too wide. 
Models with reacceleration reproduce the shape of the peaks but the
absolute flux of antiprotons turns out to be too low compared to observations 
\citep{M02}.
This is why the effect considered in the present work is so important. We
investigate a self-consistent model of cosmic-ray diffusion in 
interstellar turbulence where the wave damping on energetic particles is taken
into account, and implement this effect in the GALPROP code. To make the
first calculations reasonably tractable we keep only the most essential
features of cosmic-ray diffusion in random magnetic fields and use the
simplest description of the nonlinear wave cascade in the interstellar turbulence.

\section{Equations for Cosmic Rays}

The steady state transport equation that describes diffusion and convective
transport of cosmic ray protons and nuclei in the interstellar medium is of
the form \citep[see e.g.,][for discussion]{Ber90}:
\begin{equation}
-\nabla D\nabla\Psi+\nabla(\mathbf{u}\Psi)-\frac{\partial}{\partial p}%
\left[p^{2}K\frac{\partial}{\partial p}(p^{-2}\Psi)\right]-\frac{\partial}{\partial
p}\left(  p\frac{\nabla\mathbf{u}}{3}\Psi\right)  +\frac{\partial}{\partial
p}\left(  \dot{p}_{\mathrm{loss}}\Psi\right)  +\frac{\Psi}{\tau}=q.
\label{diffeq}
\end{equation}
Here $\Psi(p,\mathbf{r})$ is the particle distribution function in momentum
$p$ normalized on total cosmic ray number density as
$N_{\mathrm{cr}}=\int dp\Psi$, $D(p,\mathbf{r})$ 
is the spatial diffusion coefficient,
$K(p,\mathbf{r})$ is the diffusion coefficient in momentum, $\mathbf{u(r)}$ is
the velocity of large-scale motions of the interstellar medium (e.g. the
velocity of a Galactic wind),$\ \dot{p}_{\mathrm{loss}}=dp/dt<0$ is the momentum
loss rate for the energetic charged particles moving through the interstellar
medium, $\tau(\mathbf{r})$ is the time scale for nuclear fragmentation. If
needed, the supplementary term which describes radioactive decay can be
added to eq.~(\ref{diffeq}). The spatial boundary condition for  $\Psi$ is
$\Psi\mid_{\Sigma}=0$,  corresponding to  the free exit of cosmic rays from
the Galaxy to intergalactic space where their density is negligible. The
region of cosmic ray diffusion is a cylinder of radius $30$ kpc and total width
 $2H$ ($H\approx4$ kpc). The source term $q(p,\mathbf{r})$ includes both
the direct production of primary energetic particles accelerated from the
thermal background in Galactic sources (e.g. supernova remnants) and the
contribution to the  nuclei considered via the processes of nuclear
fragmentation and radioactive decay of  heavier nuclei. The typical value of
the diffusion coefficient found from the fit to cosmic ray and
radioastronomical data is $D\sim3\times10^{28} $ cm$^{2}$/s at energy $\sim$1
GeV/n,  giving a diffusion mean free path $l=3D/v\sim1$ pc
($v\thickapprox c$\ is the particle velocity).

On the ``microscopic level'' the spatial and momentum diffusion of cosmic rays
results from the particle scattering on random MHD waves and discontinuities.
In the linear approximation, the Alfven (with the dispersion relation
$\omega(\mathbf{k})=k_{\parallel}V_{\mathrm{a}}$), the fast magnetosonic
($\omega(\mathbf{k})=kV_{\mathrm{a}}$), and the slow magnetosonic
($\omega(\mathbf{k})=k_{\parallel}V_{\mathrm{s}}$) waves can propagate in a
low $\beta$ plasma, $\beta=$ $(V_{\mathrm{s}}/V_{\mathrm{a}})^{2}<1$, where
$V_{\mathrm{a}}=B/\sqrt{4\pi\rho}$ is the Alfven velocity and $V_{\mathrm{s}}$
$=\sqrt{P_{\mathrm{g}}/\rho}$ is the collisionless sound velocity (for MHD
conditions it is a factor of a square root of the gas adiabatic index larger),
where $B$, $P_{\mathrm{g}}$ and $\rho$ are the magnetic field strength, the
pressure, and the mass density of the interstellar gas respectively.
In addition to waves, 
the static entropy variations can exist in the interstellar medium.
In a collisionless plasma, the slow
magnetosonic waves are not heavily damped only if the plasma is non-isothermal and the
electron temperature considerably exceeds the ion temperature. The Alfven velocity
is $1.8\times10^{5}B_{\mathrm{\mu G}}/\sqrt{n}$ cm/s, where $n$ is the
number density of hydrogen atoms. The value of the sound velocity is
$7.7\times10^{5}\sqrt{T_{4}}$ cm/s in neutral gas with temperature
$T=10^{4}T_{4}$ \textrm{K}; $V_{\mathrm{s}}$ is larger in the ionized gas
because of the free electron contribution to gas pressure $P_{\mathrm{g}}$.
The approximation of low $\beta$ plasma is valid in the dominant part of the
interstellar medium at $B\approx5$ $\mu$G since typically $n=0.002$ cm$^{-3}$
and $T=10^{6}$ K in hot HII regions, $n=0.2$ cm$^{-3}$ and $T=8\times10^{3}$ K
in warm intercloud gas, $n=30$ cm$^{-3}$ and $T=100$ K in clouds of atomic
hydrogen, $n=200$ cm$^{-3}$ and $T=10$ K in molecular clouds and thus
$\beta\approx0.1-0.3$ everywhere. The effective ``collision integral'' for
energetic charged particles moving in  small amplitude random fields $\delta
B\ll B$ can be taken from the standard quasi-linear theory of plasma
turbulence, see e.g. \citet{Ken66}. The ``collisional integral''
averaged over fast gyro-rotation of particles about the magnetic field
$\mathbf{B}$ contains the spectral densities of the effective frequencies of
particle collisions with plasmons in the form
\begin{eqnarray}
\nu_{\mu}^{\alpha}(\mathbf{k},s,p)=\frac{4\pi^{2}v}{B^{2}\left|  \mu\right|
r_{\mathrm{g}}^{2}}&&[(1-\delta_{s0})\delta(k_{\parallel}-s/(r_{\mathrm{g}
}\left|  \mu\right|  )M_{\perp}^{\alpha}(\mathbf{k})(J_{s+1}^{2}+J_{s-1}
^{2}) \nonumber \\ 
&&+\frac{\delta_{s0}}{2}\delta(k_{\parallel}-k_{\perp}V_{\mathrm{a}}/(v\left|
\mu\right|  ))J_{1}^{2}M_{\parallel}^{\alpha}(\mathbf{k})k_{\parallel}
^{2}k_{\perp}^{-2}], 
\label{eq2}
\end{eqnarray}
\newline $k_{\parallel}>0$ \citep[as presented by][]{Ptu89}.
Here the waves are assumed symmetric about the
magnetic field direction and to have zero average helicity.
The index $\alpha$ characterizes the type of
waves including the direction of their propagation, $\mu$ is the cosine of
pitch angle, the energy densities of random magnetic fields for perpendicular
and parallel  to the average field $\mathbf{B}$ components are
$M_{\perp}^{\alpha}(\mathbf{k})$ and $M_{\parallel}^{\alpha}(\mathbf{k})$
respectively, $J_{m}=J_{m}(k_{\perp}r_{g}\sqrt{1-\mu^{2}})$ is the Bessel
function. It is explicitly taken into account that the scattering at $s=0$
occurs only on the fast magnetosonic waves propagating almost perpendicular to the
magnetic field. The Larmor radius is $r_{\mathrm{g}}=pc/(ZeB)=3.3\times
10^{12}R_{\mathrm{GV}}/B_{\mathrm{\mu G}}$ cm, where the particle magnetic
rigidity $R=pc/Ze$ is measured in GV, and the average magnetic field is
measured in $\mathrm{\mu G}$. The wave-particle interaction is of resonant
character so that an energetic particle is predominantly scattered by the
irregularities of magnetic field $\delta B$ that have the projection of wave
vector on the magnetic field direction equal to $k_{\parallel}=\pm s/\left(
r_{\mathrm{g}}\mu\right)  $. The integers  $s=0,1,2...$ correspond to
the cyclotron resonances of different orders. The efficiency of particle
scattering depends on the polarization of the waves and on their distribution in
$\mathbf{k}$-space. The first-order resonance $s=1$\ is the most important for
the isotropic and also for the one-dimensional distribution of random MHD
waves along the average magnetic field, $\mathbf{k}\parallel\mathbf{B}$. In
some cases -- for calculation of scattering at small $\mu$ and for
calculation of perpendicular diffusion -- the broadening of resonances and
magnetic mirroring effects should be taken into account.

The evolution of the particle distribution function on time-scales
$\triangle t\gg\nu^{-1}$ and distances $\triangle z\gg v\nu^{-1}$ can be
described in the diffusion approximation with the following expressions for
the spatial diffusion coefficient along the magnetic field $D_{\parallel}$ and
the diffusion coefficient in momentum $D_{pp}$:
\begin{equation}
D_{\parallel}(p)=\frac{v^{2}}{4}\int_{-1}^{+1}d\mu(1-\mu^{2})\left(
{\textstyle\sum_{\alpha,s,\mathbf{k}}}
\nu_{\mu}^{\alpha}(\mathbf{k},s,p)\right)  ^{-1}, 
\label{eq3}
\end{equation}
\begin{equation}
D_{pp}(p)  
=\frac{p^{2}}{4}\int_{-1}^{+1}d\mu(1-\mu^{2})
\left[
{\textstyle\sum_{\alpha,s,\mathbf{k}}}
\nu_{\mu}^{\alpha}(\mathbf{k},s,p)\frac{\left[V_{f}^{\alpha}(\mathbf{k})\right]^{2}%
}{v^{2}}-\left(
{\textstyle\sum_{\alpha,s,\mathbf{k}}}
\nu_{\mu}^{\alpha}(\mathbf{k},s,p)\frac{V_{f}^{\alpha}(\mathbf{k})}{v}\right)
^{2}\left(
{\textstyle\sum_{\alpha,s,\mathbf{k}}}
\nu_{\mu}^{\alpha}(\mathbf{k},s,p)\right)  ^{-1}\right]  \!,
\label{eq4}
\end{equation}
see \citet{Ber90} where the analogous equations were derived for
one-dimensional turbulence with $k_{\perp}=0$. Here $V_{\mathrm{f}}^{\alpha}(\mathbf{k}
)=\omega^{\alpha}(\mathbf{k})/k_{\parallel}$, the summations contain integrals
over $\mathbf{k}$-space, and the terms which contain $\nu_{\mu}^{\alpha
}(\mathbf{k},s=0,p)$ should be corrected in eqs.~(\ref{eq3}), (\ref{eq4}) 
compared to eq.~(\ref{eq2}):
multiplied by $(1-\mu^{2})^{2}$ in $\left(
{\textstyle\sum}
\nu_{\mu}^{\alpha}\right)  ^{-1}$, and multiplied by $(1-\mu^{2})$ in other
terms. One can check that the interaction of energetic particles with slow
magnetosonic waves is relatively weak (as $V_{\mathrm{s}}^{2}/V_{\mathrm{a}
}^{2}\ll1$) and can be ignored.

Locally, the cosmic ray diffusion is anisotropic and occurs along the local
magnetic field because the particles are strongly magnetized, $r_{\mathrm{g}
}\ll l$. The isotropization is accounted for by the presence of strong
large-scale ($\sim$100 pc) fluctuations of the Galactic magnetic field. The
problem is not trivial even in the case of relatively weak large scale random
fields, since the field is almost static and the strictly one-dimensional
diffusion along the magnetic field lines does not lead to non-zero diffusion
perpendicular to $\mathbf{B}$, see \citet{Chu93}, \citet{Gia99}, \citet{Cas01}.

The eqs.~(\ref{eq3}) and (\ref{eq4}) are too cumbersome for our present application.
Based on eq.~(\ref{eq3}) and with  reference to the detailed treatment of cosmic ray
diffusion by \citet{Top85} and \citet{Ber90}, we use below the
following simplified equation for the diffusion coefficient:
\begin{equation}
D=vr_{\mathrm{g}}B^{2}/\left[  12\pi k_{\mathrm{res}}W(k_{\mathrm{res}
})\right], 
\label{eq5}
\end{equation}
where $k_{\mathrm{res}}=1/r_{g}$ is the resonant wave number, and $W(k)$ is
the spectral energy density of waves normalized as $\int dkW(k)=\delta
B^{2}/4\pi$. The random field at the resonance scale is assumed to be weak,
$\delta B_{\mathrm{res}}\ll B$.

The cosmic ray diffusion coefficient is equal \ to
\begin{equation}
D=\frac{vr_{\mathrm{g}}^{a}}{3(1-a)k_{\mathrm{L}}^{1-a}}\frac{B^{2}}{\delta
B_{\mathrm{L}}^{2}} 
\label{eq6}
\end{equation}
for particles with $r_{\mathrm{g}}<k_{\mathrm{L}}^{-1}$\ under the assumption
of a power law spectrum of turbulence $W(k)\propto1/k^{2-a},$
$k>k_{_{\mathrm{L}}}$. We introduced here the principle wave number of the
turbulence $k_{L}$ and the amplitude of random field $\delta B_{L}$
at this scale. The diffusion has scaling $D\propto v(p/Z)^a$.

The diffusion in momentum is described by the following equation which is a
simple approximation of eq.~(\ref{eq4}):
\begin{equation}
K=p^{2}V_{\mathrm{a}}^{2}/\left(9D\right). 
\label{eq7}
\end{equation}

Eqs.~(\ref{eq5})-(\ref{eq7}) give estimates of particle diffusion in position and momentum
needed in eq.~(\ref{diffeq}). They reflect the most essential features of cosmic-ray
transport: the frequency of particle scattering on random magnetic field is
determined by the energy density of this field at the resonance wave number
$k_{\mathrm{res}}$; the acceleration is produced by the waves moving with
typical velocity $V_{\mathrm{a}}$ and is stochastic in nature. Eq.~(\ref{eq5}) implies
equal intensities of waves moving along the magnetic field in opposite
directions, the imbalanced part of the total wave energy density should not be 
taken into account when calculating $K(p)$, see \citet{Ber90} for
details. Eqs.~(\ref{eq5})-(\ref{eq7}) are also valid for small-amplitude nonlinear waves
including weak shocks. The equations can be used for the isotropic distribution of
Alfven and fast magnetosonic waves and they give correct order of magnitude
estimates for the wave distribution concentrated around the direction of
average magnetic field. It should be pointed out however that the
isotropization of the diffusion tensor does not occur in the case of a pure
parallel propagation of waves ($\mathbf{k}\parallel\mathbf{B}$). Another
special case is  $2$D turbulence with perpendicular propagation of waves
($\mathbf{k}\perp\mathbf{B}$).  In this case, the scattering occurs only for
magnetosonic waves through the resonance $s=0$ which leads to a very large
diffusion coefficient, about a factor $(v/V_{\mathrm{a}})^{2}$ larger than
given by eqs.~(\ref{eq5}), (\ref{eq6}).

Eq.~(\ref{eq6}) shows that the level of interstellar turbulence that is needed to
account for the diffusion of GeV cosmic rays is very small: $\delta
B_{\mathrm{res}}^{2}/B^{2}=(1-a)r_{\mathrm{g}}/l\sim10^{-6}$ at
$k_{\mathrm{res}}^{-1}\sim10^{12}$ cm (if $a\sim0.5$). An extension to smaller
wave numbers gives $\delta B^{2}(>k)/B^{2}\sim10^{-6}(10^{12}k)^{1-a}$ where
$k$ is in cm$^{-1}$.

\section{Equations for Interstellar Turbulence}

The description of MHD turbulence is a complicated and not completely solved
problem even in the case of small-amplitude random fields. Comprehensive
reviews of  MHD turbulence have been given by \citet{Ver04}, and by
\citet{Zho04}, \citet{Elm04}, and \citet{Sca04} with
application to  interstellar turbulence.

The classic problem is the determination of the wave spectrum in the presence of
sources at small wave numbers $k\sim k_{\mathrm{L}}$ and the strong dissipation
at much larger wave numbers (in some cases the cascade is inverse). Note that
the spectrum of interstellar MHD turbulence determines the transport
coefficients eqs.~(\ref{eq5}), (\ref{eq7}). According to the Kolmogorov-Obukhov hypothesis
\citep{Kol41,Obu41}, the resulting spectrum at intermediate $k$,
i. e. in the inertial range, is characterized by a constant energy flux to
higher wave numbers. The hypothesis was originally suggested for the
description of developed hydrodynamic turbulence in incompressible fluids. The
Kolmogorov spectrum is of the form $W(k)\varpropto k^{-5/3}$. The spectrum of
weak acoustic turbulence $W(k)\varpropto k^{-3/2}$ was found by
\citet{Zak70}. This result was criticized by
\citet{Kad73}. They argued that the developed shocks produce an additional
dissipation of acoustic wave energy. The shock-dominated turbulence (also
called the Burgers turbulence) is characterized by the spectrum $k^{-2}$
irrespective of the nature of dissipation at the shock front.

The presence of magnetic field in MHD turbulence complicates the issue because
the turbulence becomes anisotropic and new types of waves arise in a
magnetized medium. Generally, all gradients are larger perpendicular to the
field and all perturbations are elongated along the magnetic field direction
even with isotropic excitation. \citet{Iro63} and \citet{Kra65} gave
the first phenomenological theory of MHD turbulence and obtained the spectrum
$W(k)\varpropto k^{-3/2}$. 
 They assumed that small-scale fluctuations are
isotropic, which  is contradictory because wave interactions break the isotropy
in the presence of an external magnetic field.

Direct observations of MHD turbulence in
the solar wind plasma where $\beta\sim1$ have shown the existence of a
Kolmogorov-type turbulence spectrum that contains waves moving both along the
magnetic field and in almost perpendicular directions \citep{Sau99}.
Such a spectrum was obtained over several decades of wave numbers in solar
wind radio propagation studies \citep{Woo79}.  Numerical
simulations of incompressible ($\beta\gg1$) MHD turbulence favored the
Kolmogorov spectrum \citep{Ver96}.

Over the past decade, there has been a renewed interest in understanding of
magnetohydrodynamic turbulence as it applies to interstellar magnetic field
and density fluctuations (\citealt{Gol95,Gol97,Ng97,Gal00}, see also earlier work by
\citealt{She83}). \citet{Gol95} exploited anisotropy in MHD turbulence
and obtained Kolmogorov-like spectrum for the energy density of Alfven waves. The
``elementary interactions'' between Alfven waves satisfies the three-wave
resonance conditions. However there is no exact relation between wave number
and frequency in this case of strong turbulence. The main part of the energy
density in this turbulence is concentrated  perpendicular to the local
magnetic field wave vectors $k_{\perp}\approx k$, while the parallel wave
numbers are small: $k_{\parallel}\sim\left[  kW(k) /\left( B_{0}^{2}/4\pi
\right)\right]  ^{1/2}k_{\perp}$. The cascade is anisotropic with
energy confined within the cone $k_{\parallel}\varpropto k_{\perp}^{2/3}$.
Numerical simulations have confirmed this concept (\citealt{Cho00}). 

 Although the formalism has been developed
for incompressible MHD turbulence, 
\citet{Lit01} argued that the compressibility does not
essentially alter the results on the Alfven wave spectrum. The distribution of
slow magnetosonic waves passively follows that of Alfven waves. The fast
magnetosonic waves have an independent nonlinear cascade which is isotropic and
has a Kraichnan-type spectrum $W(k)\varpropto k^{-3/2}$. These conclusions were
supported by numerical simulations by \citet{Cho02}.

The  description of weak MHD turbulence in low $\beta$ plasma outlined above is
probably not complete and needs further analysis before it is accepted as a
standard model of interstellar turbulence. First, there is still the discrepancy
between theoretical results of different authors. Thus considering the
scattering of Alfven waves and fast magnetosonic waves on slow magnetosonic
waves, \citet{Kuz01} found the Kraichnan-type spectra for all these types
of waves with their preferentially parallel propagation, which  disagrees with
\citet{Lit01}. Second,  consideration of processes in
turbulent collisionless plasmas at the kinetic level involves additional
nonlinear processes of induced wave scattering on thermal ions \citep{Liv70} 
that may change the spectra \citep{Cha85}.
Third, the real turbulence can be strongly intermittent, imbalanced, etc.
\citep[e.g.][]{Lit03}, which  may also affect the interstellar MHD spectrum.

Information on the extended interstellar turbulence spectrum has been obtained from 
radio scintillation and refraction observations (sensitive to fluctuations of
thermal electron density),  measurements of the differential Faraday
rotation angles from distant sources (mainly produced by fluctuations in the
interstellar magnetic field), and the observations of random motions in the
interstellar gas. These data are consistent with the assumption that a single
close-to-Kolmogorov spectrum extends from scales $10^{8}$ to $10^{20}$ cm
\citep{Lee76}, see \citet{Arm81,Arm95} and references therein.

In the absence of an easily manageable and commonly accepted exact equation for the
energy density $W(k)$, we employ below the simplest steady state
phenomenological equation that represents the concept of a wave cascade in the
inertial range of wave numbers:
\begin{equation}
\frac{\partial}{\partial k}\left(  \frac{kW(k)}{T_{\mathrm{nl}}}\right)  =0
\label{eq8}
\end{equation}
(\citealt{Tu88,Nor96}; this approach goes back to
\citealt{Cha48} and \citealt{Hei48}). Here the approximation of a characteristic time
$T_{\mathrm{nl}}$ is used to account for the non-linear wave interactions
which provide the transfer of energy in k-space\footnote{See also
\citet{Cra03} and references therein for possible modifications of
eq.~(\ref{eq8}) when a diffusive flux of wave energy
in $k$-space is assumed and when the equation is written
for a power spectrum $W(\mathbf{k})$  not averaged over direction
(here $d^{3}kW(\mathbf{k})=dkW(k)$).}.

Formally, the Kolmogorov-type spectrum $W(k)\varpropto k^{-5/3}$ follows from
eq.~(\ref{eq8}) if $T_{\mathrm{nl}}=T_{\mathrm{A}}=$ $\left[  C_{\mathrm{A}}
k\sqrt{kW(k)/\left(  4\pi\rho\right)}\right]  ^{-1}$, where the constant
$C_{\mathrm{A}}\sim0.3$ as can be estimated from the simulations of Verma et
al (1996). The Kraichnan spectrum $W(k)\varpropto k^{-5/3}$ is obtained from
eq.~(\ref{eq8}) if $T_{\mathrm{nl}}=T_{\mathrm{M}}=$ $\left[  C_{\mathrm{M}}
k^{2}W(k)/\left(  \rho V_{a}\right)\right]  ^{-1}$, where $C_{\mathrm{M}
}\sim1$.

The Kolmogorov-type nonlinear rate $1/T_{\mathrm{A}}$ is relatively high since
it is proportional to the amplitude of weak random field whereas the Kraichnan
rate $1/T_{\mathrm{M}}$ is proportional to the amplitude squared. Our
preliminary estimates \citep{Mos03,Ptu03} showed
that the MHD cascade with the Kolmogorov rate is not significantly affected by
damping on cosmic rays even if the waves propagate along the magnetic field
which  makes the wave-particle interactions the most efficient. In addition, if
the concept of \citet{Gol95} works for interstellar turbulence,
then the Kolmogorov rate refers to the Alfven waves which are distributed
almost perpendicular to an external magnetic field and thus do not produce any
significant scattering of cosmic rays, see Section 2.\ So, we do not consider
the damping of cascades with the Kolmogorov rate in what follows.

For a cascade with the Kraichnan rate $1/T_{\mathrm{M}}$ which describes an
isotropic or close to parallel distribution of MHD waves
\citep[according to][it refers only to fast magnetosonic waves but
not to Alfven waves]{Gol95}, the equation for wave energy density can be written 
as follows:

\begin{equation}
\frac{\partial}{\partial k}\left(  \frac{C_{\mathrm{M}}}{\rho V_{\mathrm{a}}
}k^{3}W^{2}(k)\right)  =-2\Gamma(k)W(k)+S\delta(k-k_{\mathrm{L}}), 
\label{eq9}
\end{equation}
$k\geq k_{\mathrm{L}}$. Here $\Gamma=\Gamma_{\mathrm{cr}}+\Gamma_{\mathrm{th}
}$ is the wave attenuation rate on cosmic-ray particles ($\Gamma_{\mathrm{cr}
}$) and on thermal particles ($\Gamma_{\mathrm{th}}$), $S$ characterizes the
source strength. The main sources of interstellar turbulence are supernova
explosions, winds of massive stars, superbubbles, and differential rotation of the
Galactic disk. These sources produce strong random magnetic fields and probably
initiate nonlinear wave cascades at the scale $k_{\mathrm{L}}^{-1}\sim100$ pc.

Below we ignore the dissipation on thermal particles (see discussion below)
and set $\Gamma_{\mathrm{th}}=0$ to study purely the effect of damping on cosmic
rays. We actually deal with a small part of the possible global wave spectrum at
wave numbers $k\gtrsim10^{-14}$ cm$^{-1}$ which  resonantly scatter cosmic rays
with energies less than about $100$ GeV/n. The existence of a single
interstellar MHD spectrum over $12$ orders of magnitude is an unsolved problem
in itself and  is beyond the scope of the present work. The quantities
$k_{\mathrm{L}}$ and $W(k_{\mathrm{L}})$ are used here only for purposes of normalization.

The equation for  wave amplitude attenuation on cosmic rays is \citep{Ber90}:
\begin{equation}
\Gamma_{\mathrm{cr}}(k)=\frac{\pi Z^{2}e^{2}V_{\mathrm{a}}^{2}}{2kc^{2}}
\int_{p_{\mathrm{res}}(k)}^{\infty}\frac{dp}{\,p}\Psi(p), 
\label{eq10}
\end{equation}
where $p_{res}(k)=ZeB/ck$.

The solution of eqs.~(\ref{eq9}), (\ref{eq10}) allows us to find the wave spectrum:
\begin{equation}
W(k)=k^{-3/2}\left[  k_{\mathrm{L}}^{3/2}W(k_{\mathrm{L}})-\frac{Z^{2}
e^{2}B^{2}V_{a}}{8C_{\mathrm{M}}c^{2}}\int_{k_{\mathrm{L}}}^{k}dk_{1}
k_{1}^{-5/2}\int_{p_{\mathrm{res}}(k_{1})}^{\infty}\frac{dp\Psi(p)}{p}\right]
, 
\label{eq11}
\end{equation}
$k>k_{\mathrm{L}}$. Here the spectral wave density at the principal scale is
determined by the source strength: $W(k_{\mathrm{L}})=\sqrt{\rho
V_{\mathrm{a}}S/C_{\mathrm{M}}}\, k_{\mathrm{L}}^{-3/2}$. The second term in
square brackets is increasing with $k$. The wave damping on cosmic rays
decreases the wave energy density and can even terminate the cascade if the
expression in square brackets reduces to zero at some $k=k_{\ast}$. The wave
energy density should be set to zero at $k>k_{\ast}$ in this case.

Now with the use of eqs.~(\ref{eq5}), (\ref{eq11}) 
one can determine the cosmic ray diffusion coefficient, which is
\begin{equation}
D(p)=\frac{D_{0}(p)}{1-g\int_{p}^{p_{_{\mathrm{L}}}}dp_{2}p_{2}^{1/2}
\int_{p_{2}}^{\infty}dp_{1}p_{1}^{-1}\Psi(p_{1})},\; 
\label{eq12}
\end{equation}%
\[
g=\frac{3\pi V_{\mathrm{a}}p^{1/2}D_{0}(p)}{2C_{\mathrm{M}}B^{2}r_{\mathrm{g}
}v}=\frac{\sqrt{Ze}B^{2}}{16\sqrt{\pi\rho c}C_{\mathrm{M}}k_{\mathrm{L}}
^{3/2}W(k_{\mathrm{L}})},
\]
$p<p_{\mathrm{L}}\mathrm{,\ }p_{\mathrm{L}}=p_{\mathrm{res}}(k_{\mathrm{L}})$.
Here $D_{0}(p)=vr_{\mathrm{g}}^{1/2}B^{2}/[12\pi k_{\mathrm{L}}^{3/2}
W(k_{\mathrm{L}})]\varpropto v(p/Z)^{1/2}$ is the diffusion coefficient
calculated for the Kraichnan-type spectrum without considering wave damping. The
second term in the denominator of eq.~(\ref{eq12}) describes the modification of the diffusion
coefficient due to  wave damping. The effect  becomes stronger to
smaller $p$. The diffusion coefficient should be formally set to infinity at
$p<p_{\ast}$ if the square bracket in eq.~(\ref{eq11}) goes to zero at some
$p=p_{\ast}$. The constant $g$ characterizes the strength of the effect for a given
cosmic ray spectrum $\Psi(p)$.

As the most abundant species, the cosmic ray protons mainly determine the
wave dissipation. Thus, with good precision only the proton component with $Z=1$
needs to be taken into account to calculate $D(p)$ by the simultaneous solution of
eqs.~(\ref{diffeq}) and (\ref{eq12}). The diffusion mean free path for other 
nuclei of charge $Z$ is $l(p/Z)$ if it is $l(p)$\ for protons.

Let us estimate the effect of wave damping. Assuming that the cosmic-ray
energy density is about $1$ eV/cm$^{3}$, the diffusion coefficient
$D=3\times10^{28}$ cm$^{2}$/s at $1$ GeV, $V_{\mathrm{a}}=10$ km/s, $B=5$
$\mu$G, and $C_{\mathrm{M}}=1$,\ the second term in the denominator in eq.~(\ref{eq12}) is
about unity at GeV energies and falls  at higher energies. We conclude
that the Kraichnan-type cascade is significantly affected by damping on cosmic
rays, and this should lead to the modification of cosmic-ray transport at
energies less than about $10$ GeV/n.

\section{Simple Selfconsistent Model}

To demonstrate the effect of wave damping, let us consider a simple case of
one-dimensional diffusion with
source distribution $q=q_{0}
(p)\delta(z)$ (corresponding to an infinitely thin disk of cosmic-ray
sources located at  the Galactic mid-plane $z=0$) and a flat cosmic-ray halo of
height $H$, see \citet{Jon01}. The source spectrum at $R<40$ GV is
$q_{0}\propto p^{-\gamma_{\mathrm{s}}}$, where  $\gamma_{\mathrm{s}}=2.5$ approximately.
Considering energetic protons, we ignore energy losses and nuclear
fragmentation ($\dot{p}_{\mathrm{loss}}=0$, $1/\tau=0$) and assume that a
Galactic wind is absent ($u=0$) in eq.~(\ref{diffeq}). Let us also assume that stochastic
reacceleration does not significantly change the energies of cosmic-ray particles
during the time of their diffusive leakage from the Galaxy and set $K=0$ in
eq.~(\ref{diffeq}). The solution of eq.~(\ref{diffeq}) in the Galactic disk is then
\begin{equation}
\Psi(p)=\frac{q_{_{0}}(p)H}{2D(p)}. 
\label{eq13}
\end{equation}

The escape length (the ``grammage''), which determines the production of
secondaries during the cosmic-ray leakage from the Galaxy, is equal to
$X=\mu_{\mathrm{g}}vH/(2D)$, where $\mu_{\mathrm{g}}$ is the gas surface
density of the Galactic disk.

The simultaneous solution of eqs.~(\ref{eq12}) and (\ref{eq13}) allows us to find $\Psi(p)$ and
$D(p)$. Introducing the function $\Psi_{0}(p)=q_{0}(p)H/[2D_{0}(p)]$, which is the
cosmic ray spectrum for an unmodified Kraichnan spectrum, and substituting
$\Psi$ from eq.~(\ref{eq13}) into eq.~(\ref{eq12}), one  finds after two differentiations
the following equation for the ratio $\varphi=D_{0}(p)/D(p)$ as a function of
$x=p^{3/2}$:
\begin{equation}
\frac{d^{2}\varphi(x)}{dx^{2}}=-\frac{4g\Psi_{0}[p(x)]}{9x}\varphi(x),
\label{eq14}%
\end{equation}
with the constraint $\varphi(\infty)=1$. Note that the same equation is valid
for the ratio $\Psi(p)/\Psi_{0}(p)$.

The function $\Psi_{0}/x$ in r.h.s.\ of eq.~(\ref{eq14}) can be approximated by a power
law function proportional to $x^{-b}$, $b=\mathrm{const}$. Thus for
$\gamma_{s}=2.5$, one has $b=3$ for ultrarelativistic ($v=c$) and $b=11/3$ for
nonrelativistic ($v\ll c$ ) protons respectively. This  allows us to find the
solution of eq.~(\ref{eq14}) in an explicit form. The diffusion coefficient
$D(p)=D_{0}(p)/\varphi(p)$ is then:
\begin{equation}
D(p)=D_{0}(p)\frac{w^{1/(b-2)}(p)}{\Gamma\left[  (b-1)/(b-2)\right]
J_{1/(b-2)}[2w(p)]},\;w(p)=\frac{2}{3(b-2)}\sqrt{gp^{3/2}\Psi_{0}(p)},
\label{eq15}
\end{equation}
where $\Gamma(z)$ is the gamma function, $J_{\nu}(z)$ is the Bessel function
of the first kind. With  $p$ decreasing from infinity and the
corresponding increase of $w(p)$  from zero, the Bessel function goes to
zero at some $w(p_{\ast})=w_{\ast}$ (where $w_{\ast}=1.92$ at $b=3$, and
$w_{\ast}=1.64$ at $b=11/3$). This means that $D(p)$ becomes infinite at
$p=p_{\ast}$ because of complete termination of the cascade; $D(p)=\infty$ at
$p\leq p_{\ast}$.

The expansion of the Bessel function for small and large arguments results in the
following approximations for eq.~(\ref{eq15}):
\begin{equation}
D(p)\approx\frac{D_{0}(p)}{1-\left\{  \Gamma\left[(b-1)/(b-2)\right]/\Gamma\left[
(2b-3)/(b-2)\right]\right\}  w^{2}(p)}\rm{\textrm{\ at }}w(p)\ll1, 
\label{eq16}
\end{equation}
and
\begin{equation}
D(p)\approx D_{0}(p)\frac{\sqrt{\pi}w^{b/[2(b-2)]}}{\Gamma\left[
(b-1)/(b-2)\right]  \sin\left[  \frac{\pi}{4}(3b-4)/(b-2)-2w(p)\right]
}\rm{\textrm{\ at }}1\lesssim w(p)<w_{\ast}. 
\label{eq17}
\end{equation}
The last approximate expression for $D(p)$ has a singularity at $w(p)=w_{\ast
}=(\pi/8)(3b-4)/(b-2)$; $w_{\ast}=1.96$ at $b=3$, and $w_{\ast
}=1.65$ at $b=11/3$ which is close to the values obtained from eq.~(\ref{eq15}). The
corresponding $p_{\ast}$ can be found from the equation $p_{\ast}^{3/2}
\Psi_{0}(p_{\ast})=\left[  3\pi(3b-4)/16\right]^{2}g^{-1}$.

\begin{figure}[tbh]
\centerline{
\includegraphics[width=3.3in]{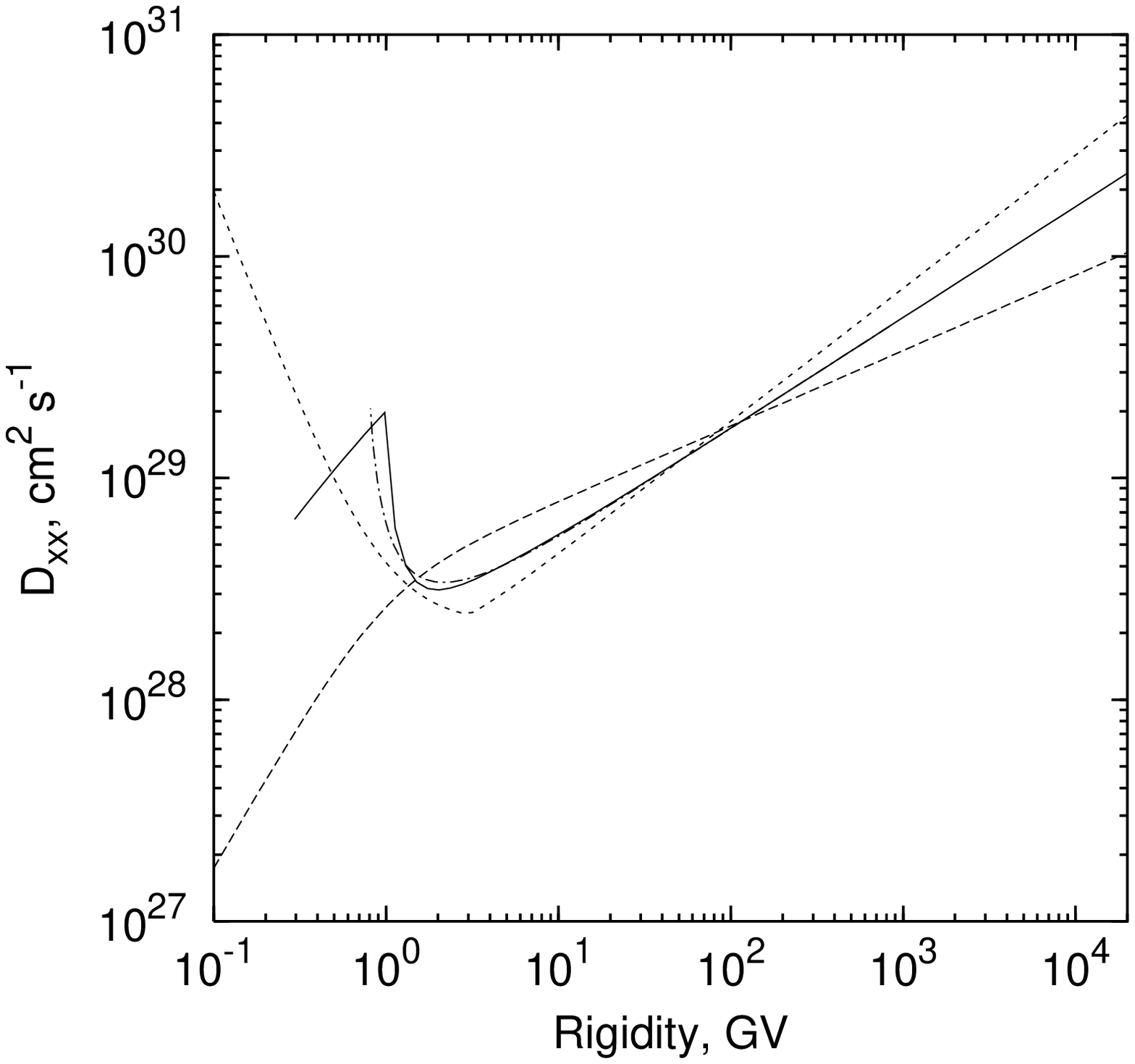}\hfill
\includegraphics[width=3.4in]{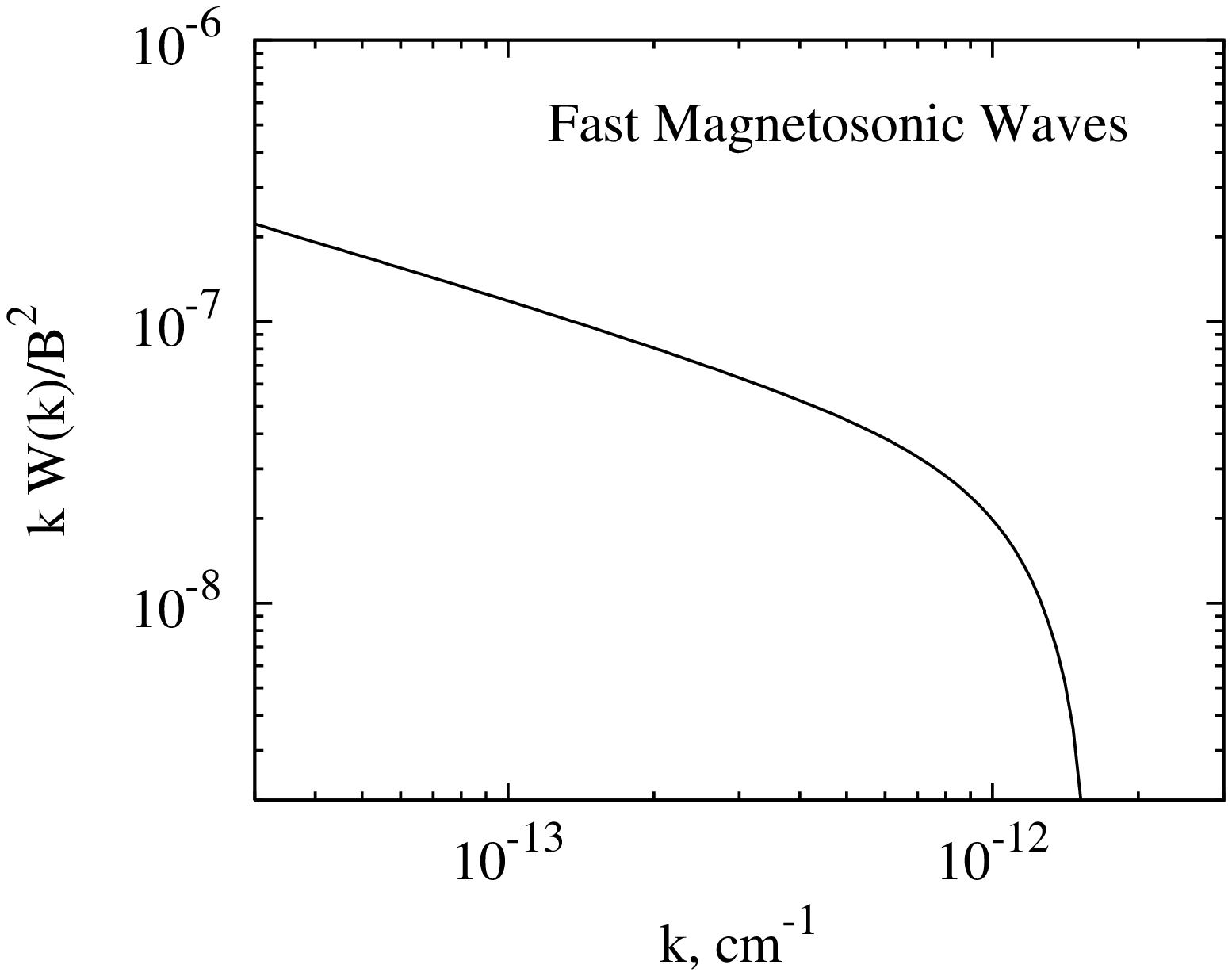}
}
\begin{minipage}[tr]{\wcap}
\caption{Diffusion coefficient $D$ for the models being discussed in this 
paper:
plain diffusion model (PD model) -- dots,
reacceleration model (RD model) -- dashes, and
diffusive reacceleration with damping model (DRD model) -- solid,
an approximate solution (eq.~[\ref{eq15}] with $b=3$) 
is shown by the dash-dot line.
\label{Dxx}}
\end{minipage} \hfill
%
\begin{minipage}[tr]{\wcap}
\caption{Dimensionless spectrum of fast
magnetosonic waves $kW(k)/B^{2}$ modified by interactions with cosmic rays
in DRD model.\label{waves}}
\vspace{2.4\baselineskip}
\end{minipage}
\vskip 1\baselineskip
\end{figure}

Eqs.~(\ref{eq15})-(\ref{eq17}) exhibit the rapid transition from unmodified to infinite
diffusion. Fig.~\ref{Dxx} demonstrates  that eq.~(\ref{eq15}) derived in the considerably
simplified model of cosmic-ray propagation (dash-dot) agrees well with the shape of the
diffusion coefficient found in the numerical simulations (solid) described in the next Section.
The results of this Section can be used for approximate estimates of cosmic ray diffusion
coefficient without invoking the full-scale calculations based on the GALPROP code.

\section{Numerical Solution with GALPROP Code}

The numerical calculations have been made using the GALPROP code for cosmic-ray
propagation. GALPROP is a flexible numerical code written in C++
which incorporates \emph{as much realistic} astrophysical input \emph{as possible}
together with latest theoretical developments. 
The model is designed
to perform cosmic-ray propagation calculations for nuclei (isotopes of H through Ni),
antiprotons, electrons and positrons, and computes \grays\ and
synchrotron emission in the same framework. The current version
of the model includes substantial optimizations in comparison
to the older model and a full 3-dimensional spatial grid for all
cosmic-ray species. It explicitly solves the full
nuclear reaction network on a spatially resolved grid.
GALPROP models have been described in detail elsewhere \citep{Str98,M02,M03}.
For the present calculation we use three-dimensional
Galactic models with cylindrical symmetry: $(R,z,p)$ -- spatial variables plus momentum.

GALPROP solves the transport equation eq.~(\ref{diffeq}) for all cosmic-ray species
using a Crank-Nicholson implicit second-order scheme.
The spatial boundary conditions assume free particle escape. 
The diffusion coefficient as a function of momentum and the reacceleration or
convection parameters are determined by boron-to-carbon (B/C) ratio data. The
spatial diffusion coefficient is taken as $D=\kappa\beta(R/R_{0})^{a}$,
$\kappa=\mathrm{const}$, $\beta=v/c$, if necessary with a break ($a=a_{1}$
below rigidity $R_{0}$, $a=a_{2}$ above rigidity $R_{0}$). The injection
spectrum of nucleons is assumed to be a power law in momentum, $q(p)\propto
p^{-\gamma_{\mathrm{s}}}$ and may also have breaks. To account for stochastic
reacceleration in the interstellar medium, the momentum-space diffusion
coefficient $K$ is related to the spatial coefficient $D$ via the Alfv\'{e}n
speed $V_{\mathrm{A}}$, see eq.~(\ref{eq7}).

The interstellar hydrogen distribution \citep{M02} uses \hi\ and CO surveys and information
on the ionized component; the helium fraction of the gas is taken as 0.11 by
number. The H$_{2}$ and \hi\ gas number densities in the Galactic plane are
defined in the form of tables, which are interpolated linearly. 
The conversion factor is taken as \Xco$\equiv
N_{\mathrm{H}_2}/W_{\mathrm{CO}}=1.9\times 10^{20}$ mols.\
cm$^{-2}$/(K km s$^{-1}$) \citep{StrongMattox96}. The extension
of the gas distribution to an arbitrary height above the plane is made using
analytical approximations. The halo size is assumed to be $H=4$ kpc. The distribution
of cosmic-ray sources is chosen to reproduce the cosmic-ray distribution
determined by analysis of EGRET \gray\  data \citep{StrongMattox96,Str98}:
\begin{equation}
\label{eq:sources}
q(R,z) = q_0 \left(\frac{R}{R_\odot}\right)^\alpha \exp\left(-\beta\frac{R-R_\odot}{R_\odot}
-\frac{|z|}{0.2{\rm\ kpc}}\right)\ ,
\end{equation}
where $q_0$ is a normalization constant, $R_\odot=8.5$ kpc, $\alpha=0.5$ and 
$\beta=1.0$ are parameters. We note that the adapted source 
distribution\footnote{A recent analysis has shown that the apparent discrepancy between 
the radial gradient in the diffuse Galactic \gray\ emissivity and the
distribution of SNR, believed to be the sources of
cosmic rays, can be plausibly solved by adopting a conversion
factor \Xco\ which increases with $R$ \citep[see a discussion in][]{S04}.}
is flatter than the distribution of SNR \citep{case98} and pulsars \citep{lorimer04}. 

\begin{figure}
\includegraphics[width=3in]{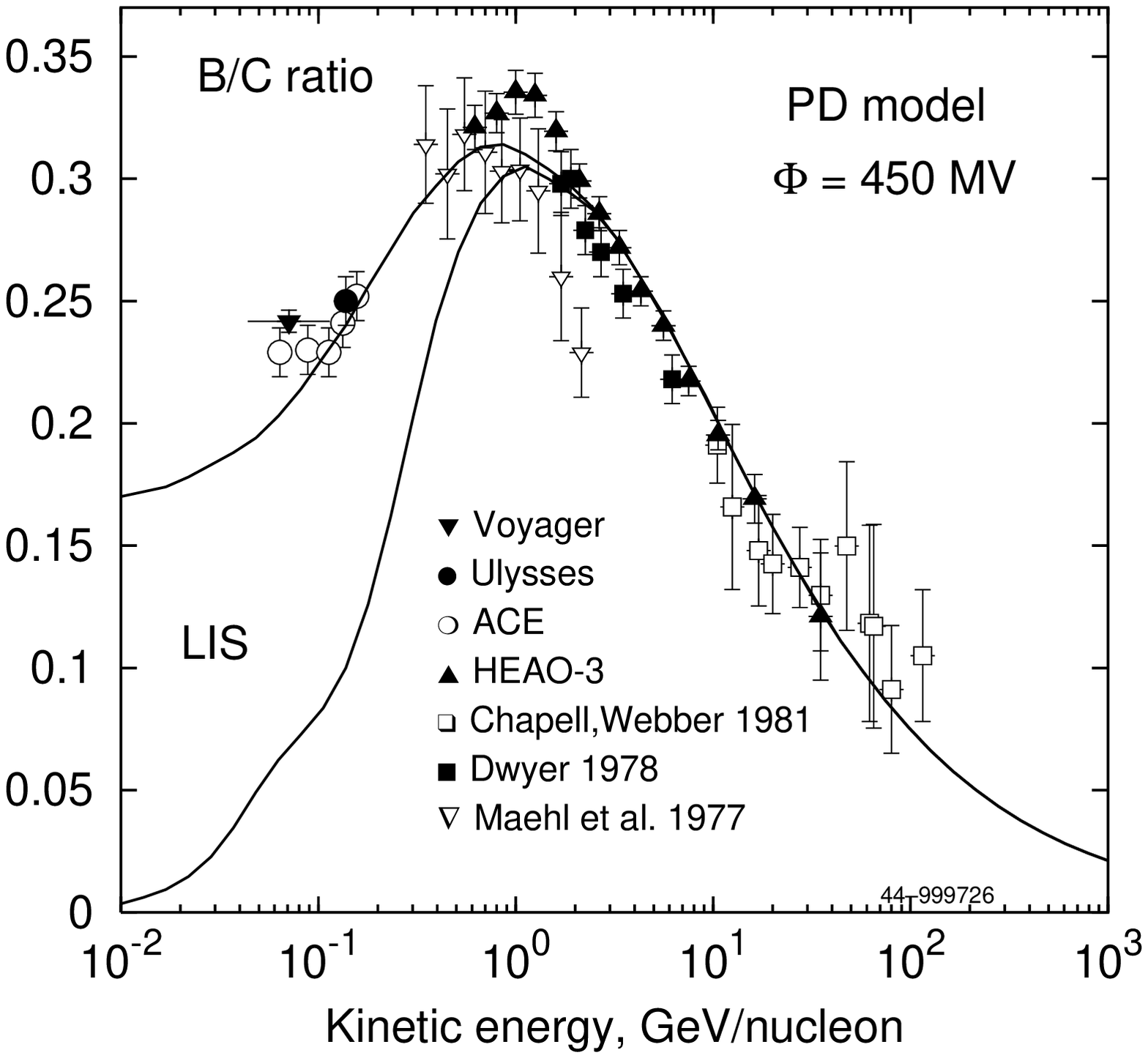}\hfill
\includegraphics[width=3in]{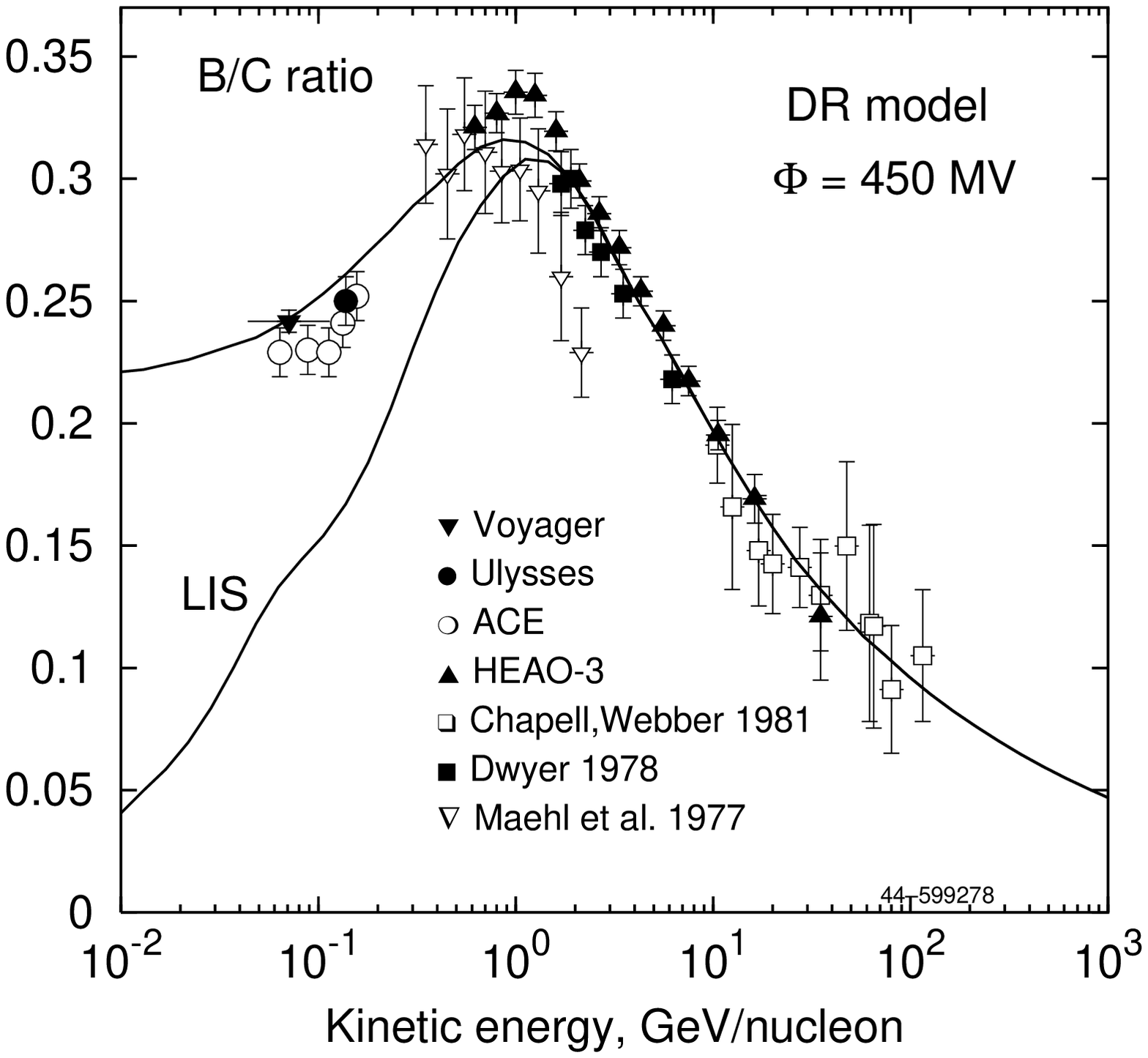}\vfill
\centerline{\includegraphics[width=3in]{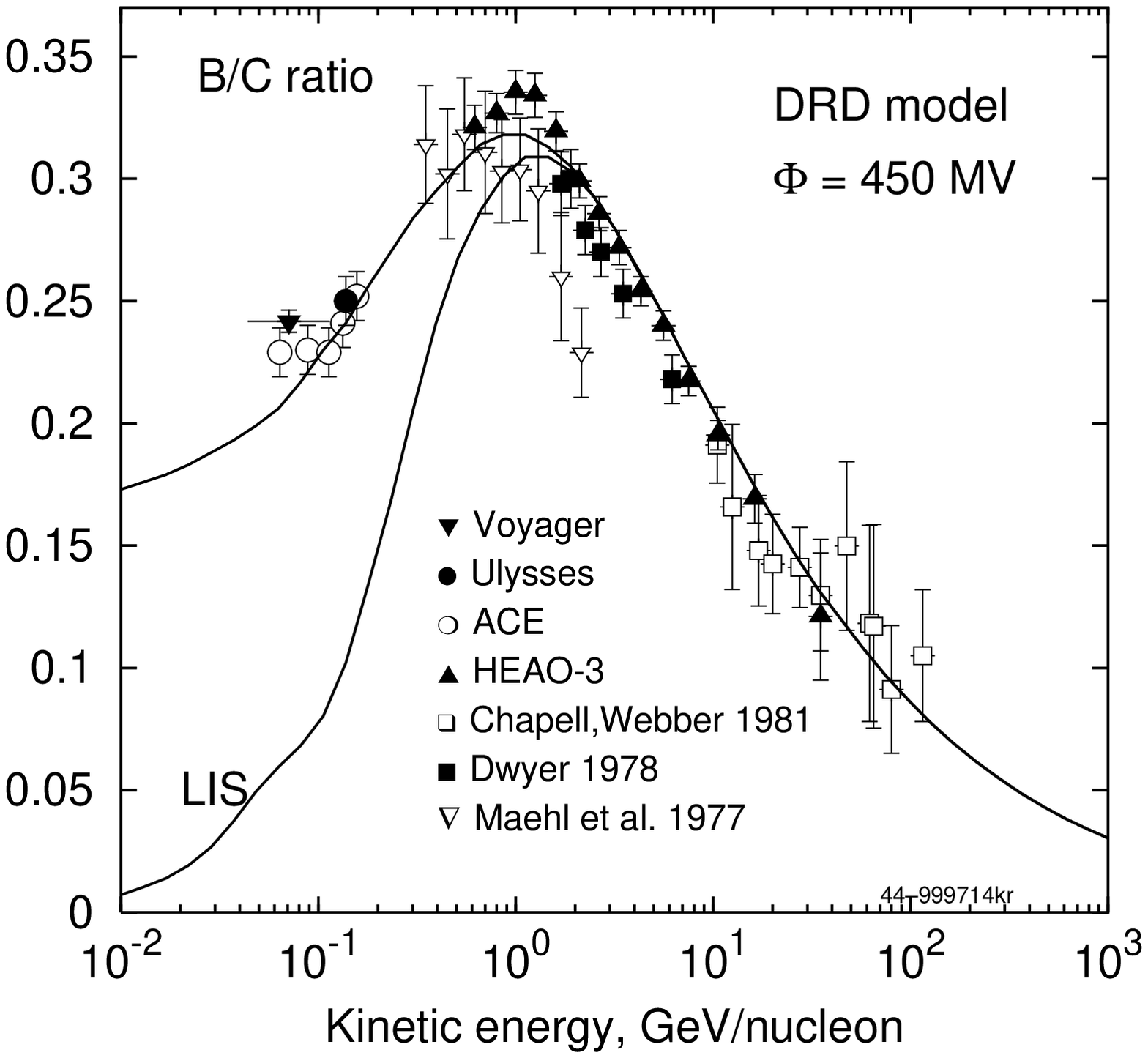}}
\caption{B/C ratio as calculated in plain diffusion model (PD model),
reacceleration model (RD model), and
diffusive reacceleration with damping model (DRD model).
Lower curve -- LIS, upper -- modulated ($\Phi=450$ MV).
Data below 200 MeV/nucleon: ACE \citep{davis}, Ulysses
\citep*{ulysses_bc}, Voyager \citep*{voyager}; high energy data:
HEAO-3 \citep{Eng90}, for other references see
\citet{StephensStreitmatter98}.\label{bc}}
\vskip 1\baselineskip
\end{figure}

The code includes cross-section
measurements and energy dependent fitting functions \citep{SM01}. The
nuclear reaction network is built using the Nuclear Data Sheets. 
The isotopic cross section database is built using the extensive
T16 Los Alamos compilation of the cross sections \citep{t16lib} and 
modern nuclear codes CEM2k and LAQGSM \citep{codes}.
The most important isotopic production cross sections 
($^2$H, $^3$H, $^3$He, Li, Be, B, Al, Cl, Sc, Ti, V, Mn) 
are calculated using our fits to major production 
channels \citep[e.g.,][]{MMS01,M03,MM03}.
Other cross sections are calculated using
phenomenological approximations by \citet{W-code} (code
\url{WNEWTR.FOR} versions of 1993 and 2003) 
and/or Silberberg and Tsao \citep{newyield} (code
\url{YIELDX\_011000.FOR} version of 2000) renormalized to the data
where it exists. For $pA$
inelastic cross section we adapted the parametrization by
\citeauthor{crosec2} \citep[code CROSEC,][]{crosec1,crosec2}.
The details of proton and antiproton cross sections are given in \citet{M02}.
Secondary positron and electron production is computed
using the formalism described in \citet{MS98}, which includes a
reevaluation of the secondary $\pi^\pm$- and $K^\pm$-meson decay
calculations. 
Energy losses for nucleons and leptons by ionization, Coulomb
scattering, bremsstrahlung, inverse Compton scattering, and synchrotron
are included in the calculations.
The heliospheric modulation is
treated using the force-field approximation \citep{forcefield}.

The reaction network is solved starting at the heaviest nuclei (i.e.,
$^{64}$Ni). The propagation equation is solved, computing all the resulting
secondary source functions, and then proceeds to the nuclei with $A-1$. The
procedure is repeated down to $A=1$. In this way all secondary, tertiary
etc.\ reactions are automatically accounted for. To be completely accurate for
all isotopes, e.g., for some rare cases of $\beta^{\pm}$-decay, the whole loop
is repeated twice.

For the present work, a  new iterative procedure which includes a
self-consistent determination of the cosmic-ray diffusion coefficient was
developed. In the first step, cosmic-ray propagation is calculated using the
undisturbed diffusion coefficient $D_{0}(p)$. In the second, we use the
propagated proton spectrum at every spatial grid point to re-calculate the
modified diffusion coefficient according to eq.~(\ref{eq12}). In the third step,
cosmic-ray propagation is calculated using the new diffusion coefficient.
Steps 2 and 3 are repeated until convergence is obtained. 
The convergence is fast and the procedure requires 
only few iterations.
In this way, using the secondary to primary ratio B/C, 
one can derive the damping constant $g=0.085$ which is considered 
as an adjustable parameter.


\begin{deluxetable}{lccccccc}
\tablecolumns{8}
\tablewidth{0pt}
\tabletypesize{\scriptsize}
\tablecaption{Propagation parameter sets.
\label{table1}}
\tablehead{
\colhead{} & 
\multicolumn{2}{c}{Injection index\tablenotemark{a}} &
\colhead{Break} &
\multicolumn{2}{c}{Diffusion coefficient @ 3 GV} & 
\colhead{Alfv\'en speed,} &
\colhead{}
\\
\cline{2-3}
\cline{5-6}
\colhead{Model} & 
\colhead{Nucleons, $\gamma_s$} &
\colhead{Electrons, $\gamma_e$} &
\colhead{rigidity, GV}&
\colhead{$\kappa$, cm$^2$ s$^{-1}$} & 
\colhead{Index, $a$} &
\colhead{$V_a$, km s$^{-1}$} &
\colhead{galdef-file} 
}
\startdata
Plain Diffusion (PD) &
2.30/2.15            &
2.40                 &
40                   &
$2.2\times10^{28}$   & 
$0.0/0.60$\tablenotemark{b} &
---                  &
44\_999726
\smallskip\\
 
Diffusive \\Reacceleration (DR)  &
1.80/2.40            &
1.60/2.50            &
4                    &
$5.2\times10^{28}$   & 
0.34                 &
36                   &
44\_599278
\smallskip\\

Diffusive \\Reacceleration with \\Damping (DRD)  &
2.40/2.24            &
2.70                 &
40                   &
$2.9\times10^{28}$   & 
0.50                 &
22                   &
44\_999714kr
\\

\enddata
\tablecomments{Adopted halo size $H=4$ kpc.}
\tablenotetext{a}{Index below/above the break rigidity.}
\tablenotetext{b}{Index below/above $R_0 = 3$ GV; $D=\beta^{-2}\kappa(R/R_0)^a$.}
\end{deluxetable}

\begin{figure}
\includegraphics[width=3in]{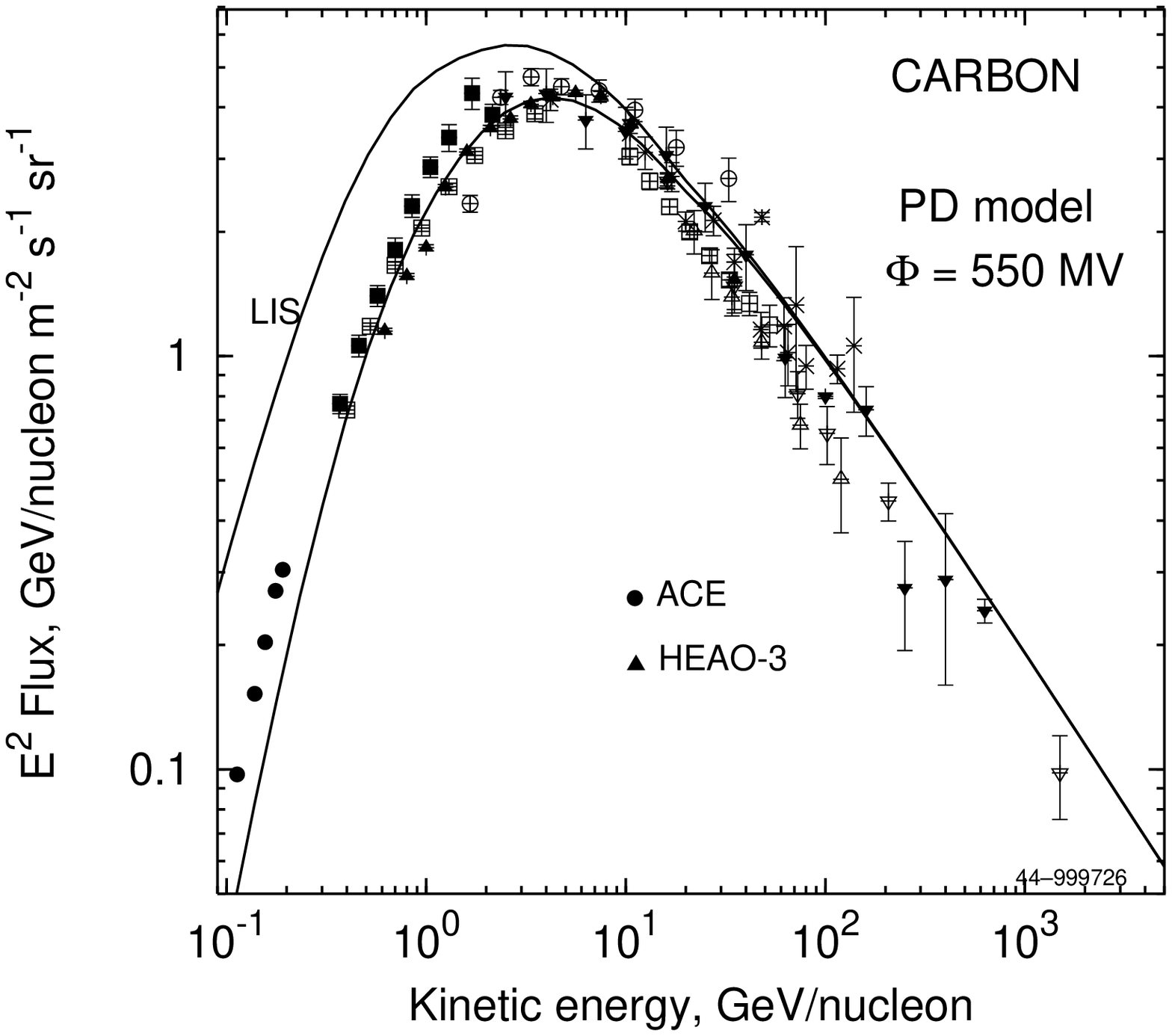}\hfill
\includegraphics[width=3in]{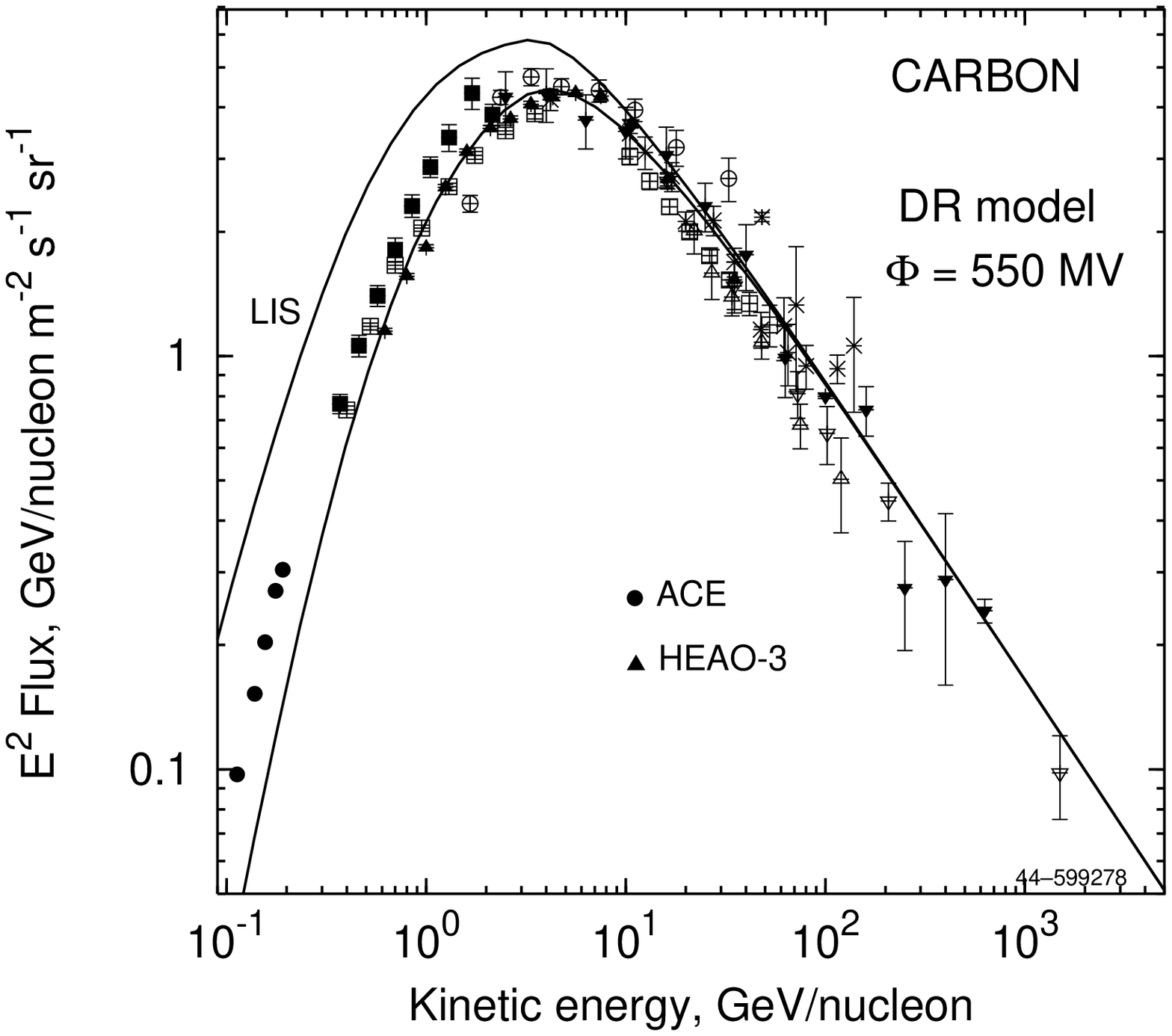}\vfill
\centerline{\includegraphics[width=3in]{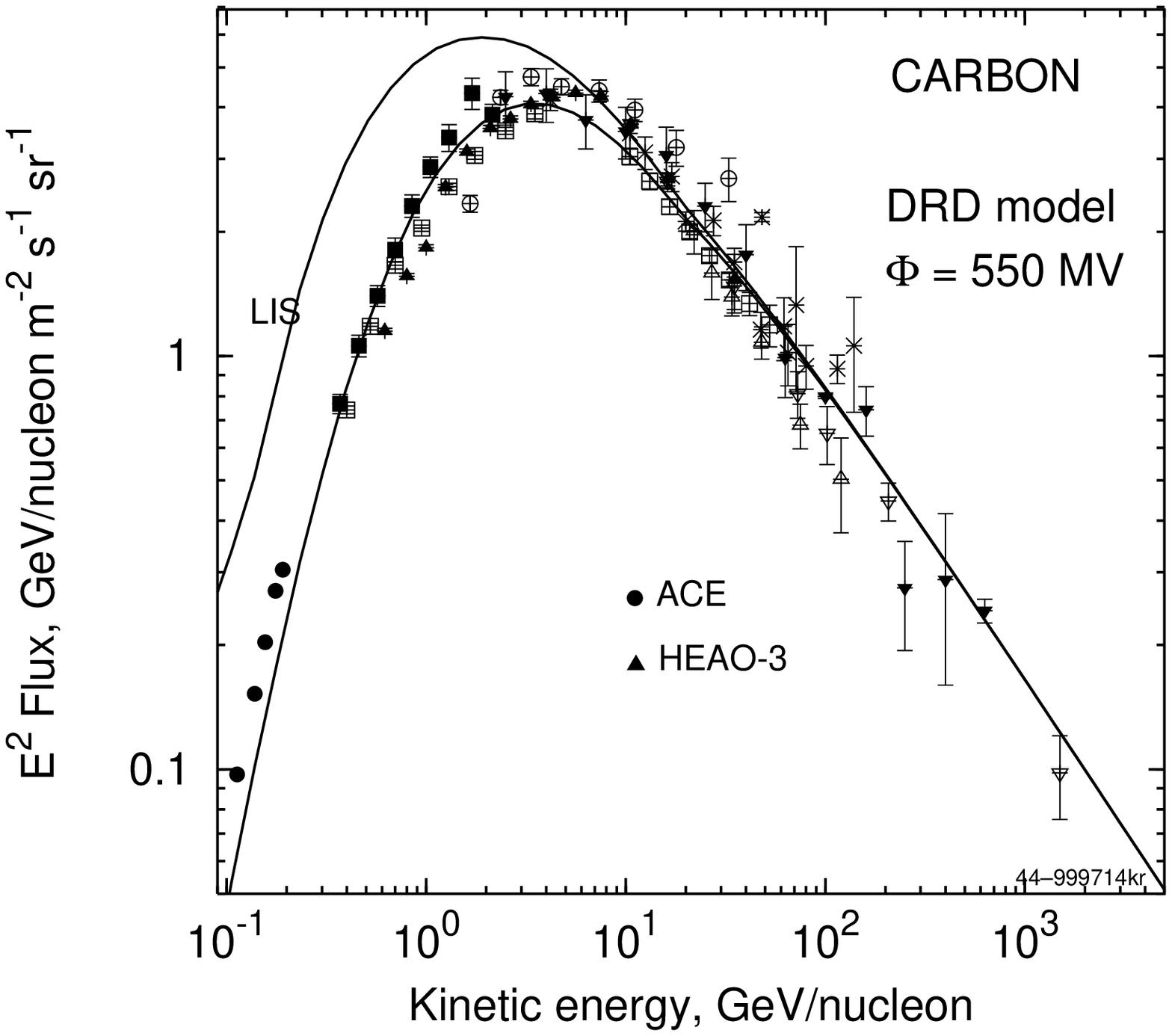}}
\caption{Spectrum of carbon calculated in plain diffusion model (PD model),
reacceleration model (RD model), and
diffusive reacceleration with damping model (DRD model). 
Upper curves -- LIS, lower curves modulated using force field
approximation ($\Phi=550$ MV). Data: ACE \citep{davis,davis01},
HEAO-3 \citep{Eng90}, for other references see
\citet{StephensStreitmatter98} (symbols are changed).\label{carbon}}
\vskip 1\baselineskip
\end{figure}

\begin{figure}
\includegraphics[width=3in]{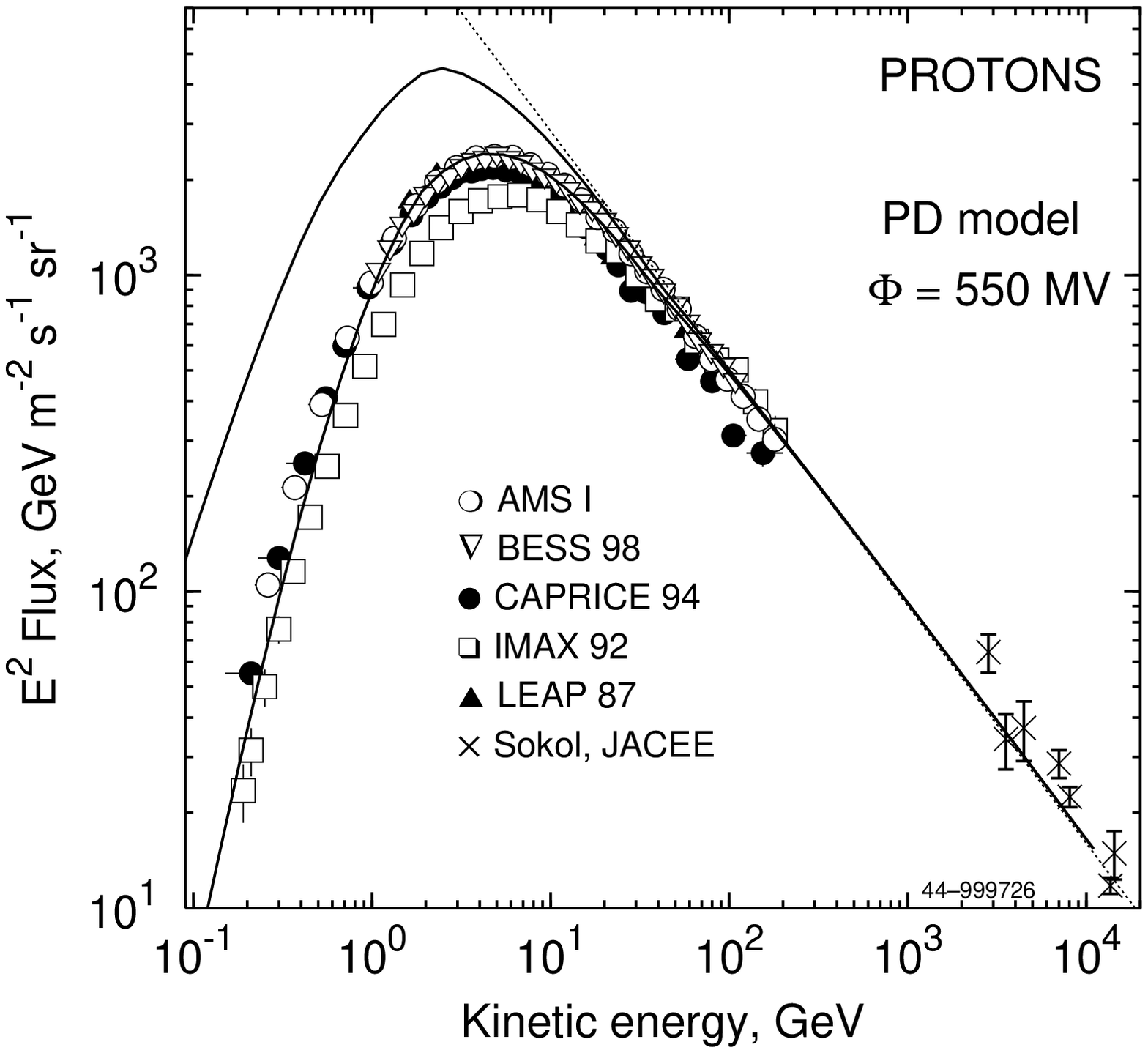}\hfill
\includegraphics[width=3in]{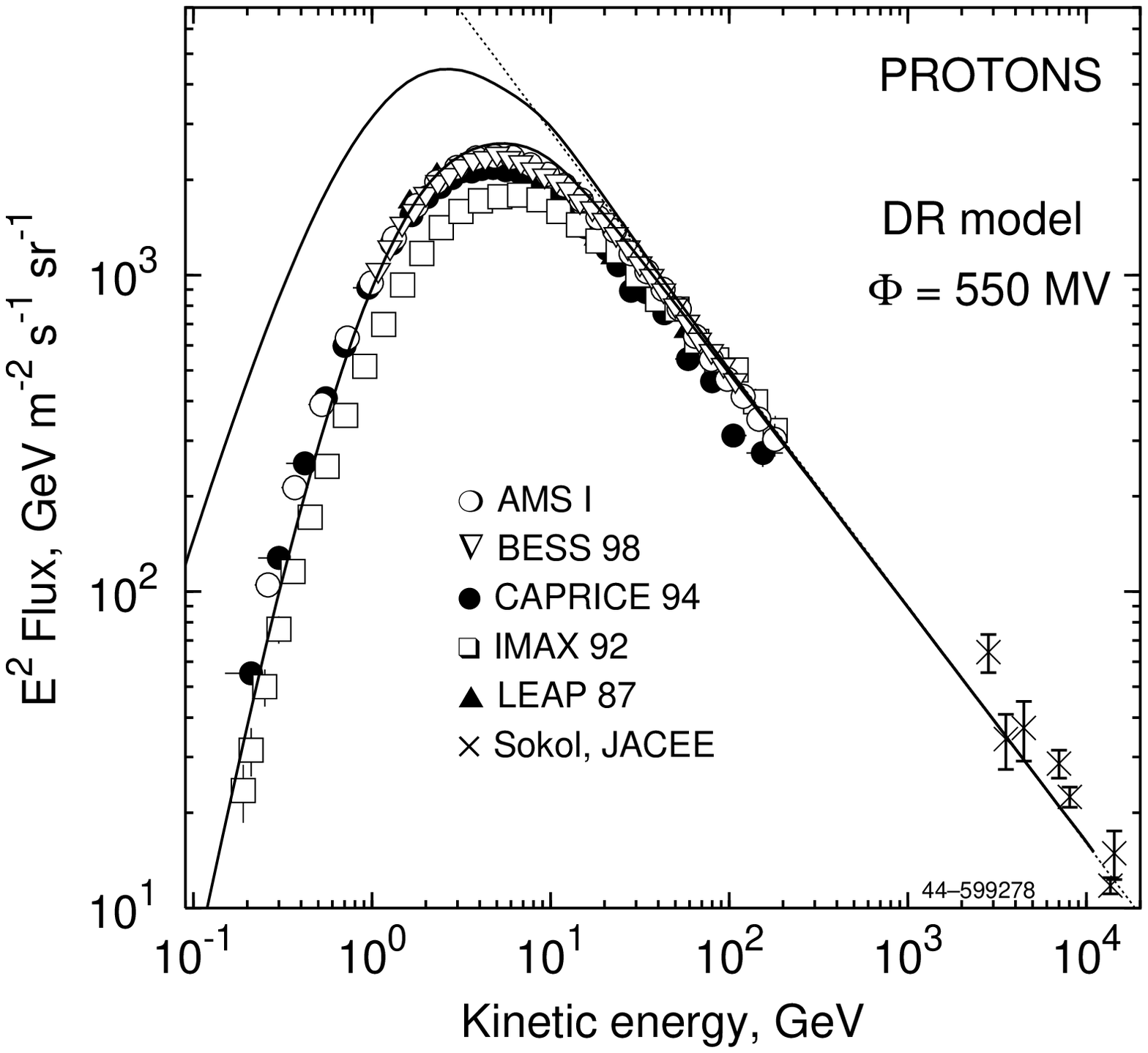}\vfill
\centerline{\includegraphics[width=3in]{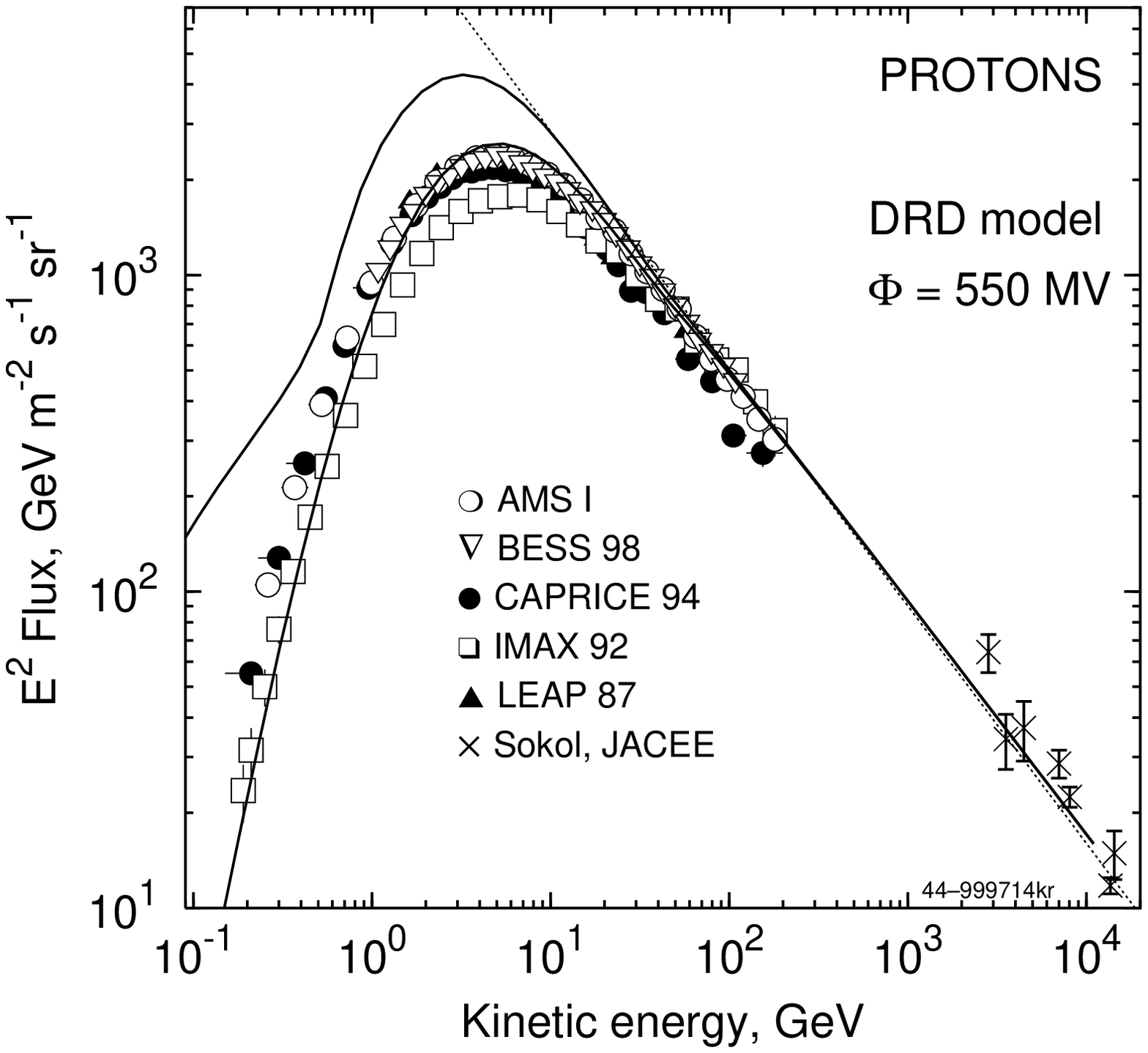}}
\caption{Proton spectra as calculated in plain diffusion model (PD model),
reacceleration model (RD model), and
diffusive reacceleration with damping model (DRD model). 
Upper curve -- LIS, lower -- modulated to 550 MV.
Thin dotted line shows the LIS spectrum best fitted to the data above 20 GeV 
\citep{M02}. 
Data: AMS \citep{p_ams}, BESS 98 \citep{sanuki00}, CAPRICE 94 \citep{Boez99},
IMAX 92 \citep{Menn00}, LEAP 87 \citep{p_leap}, Sokol \citep{sokol}, JACEE \citep{jacee}.
\label{protons}}
\vskip 1\baselineskip
\end{figure}

\begin{figure}
\includegraphics[width=3in]{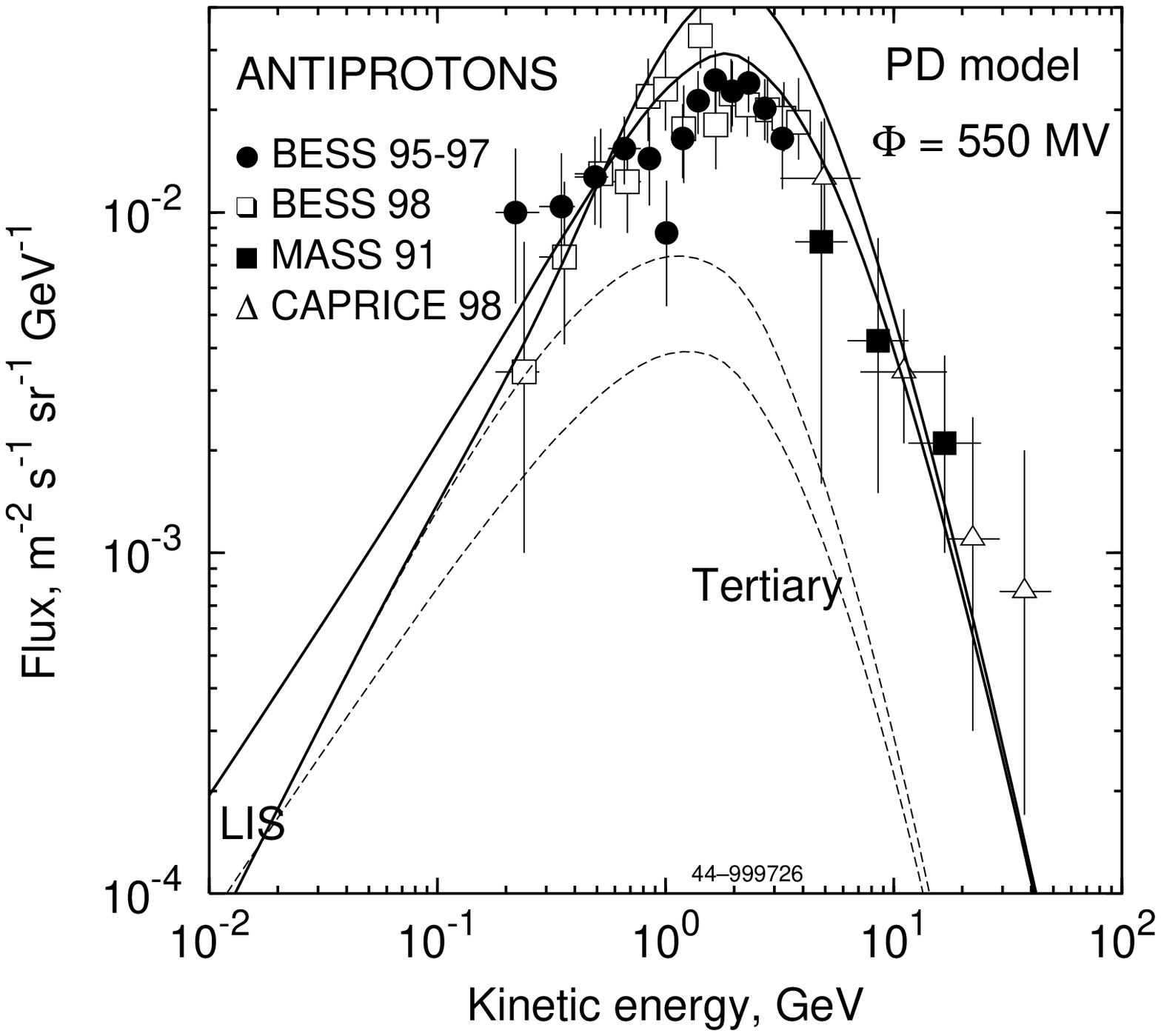}\hfill
\includegraphics[width=3in]{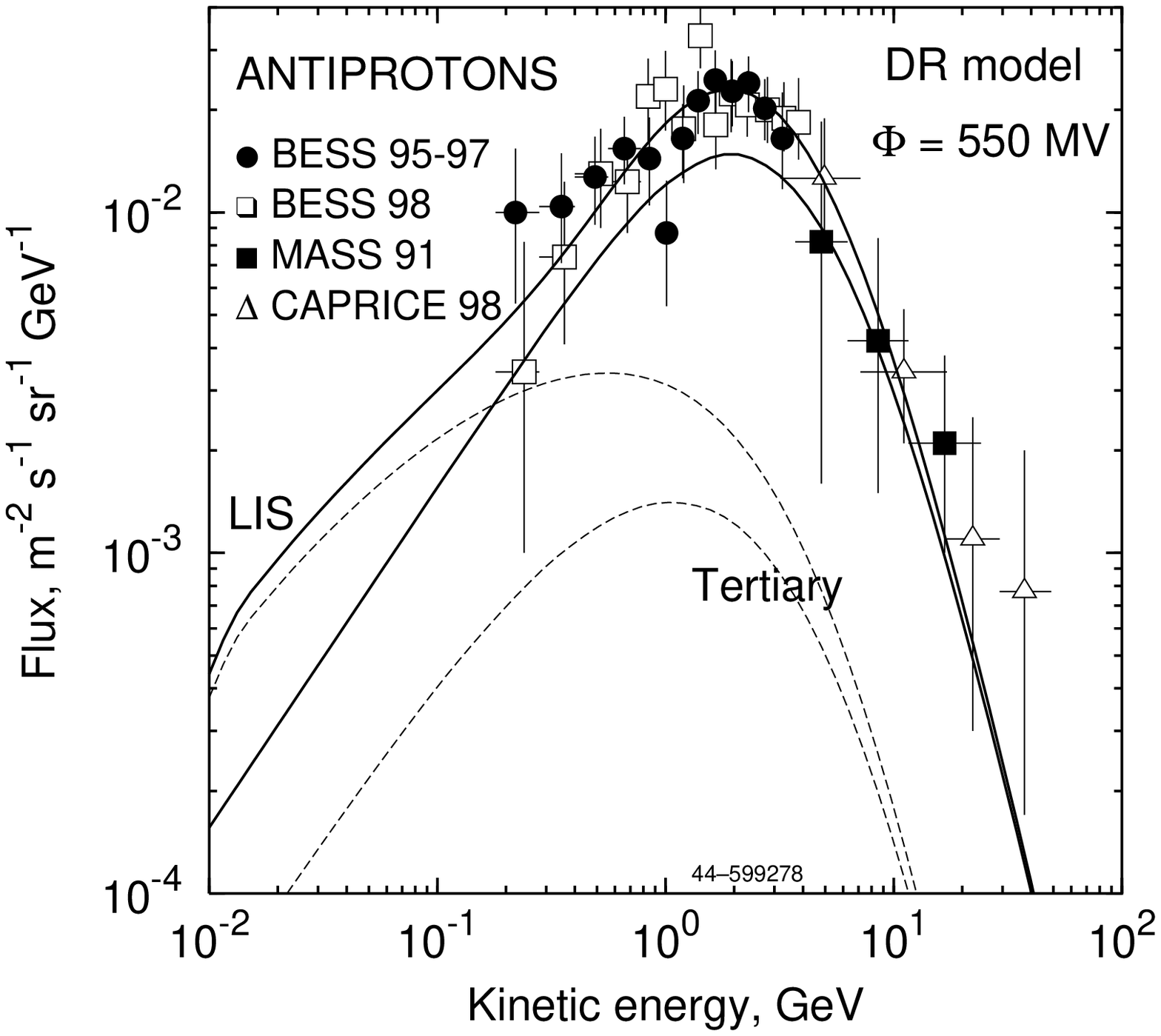}\vfill
\centerline{\includegraphics[width=3in]{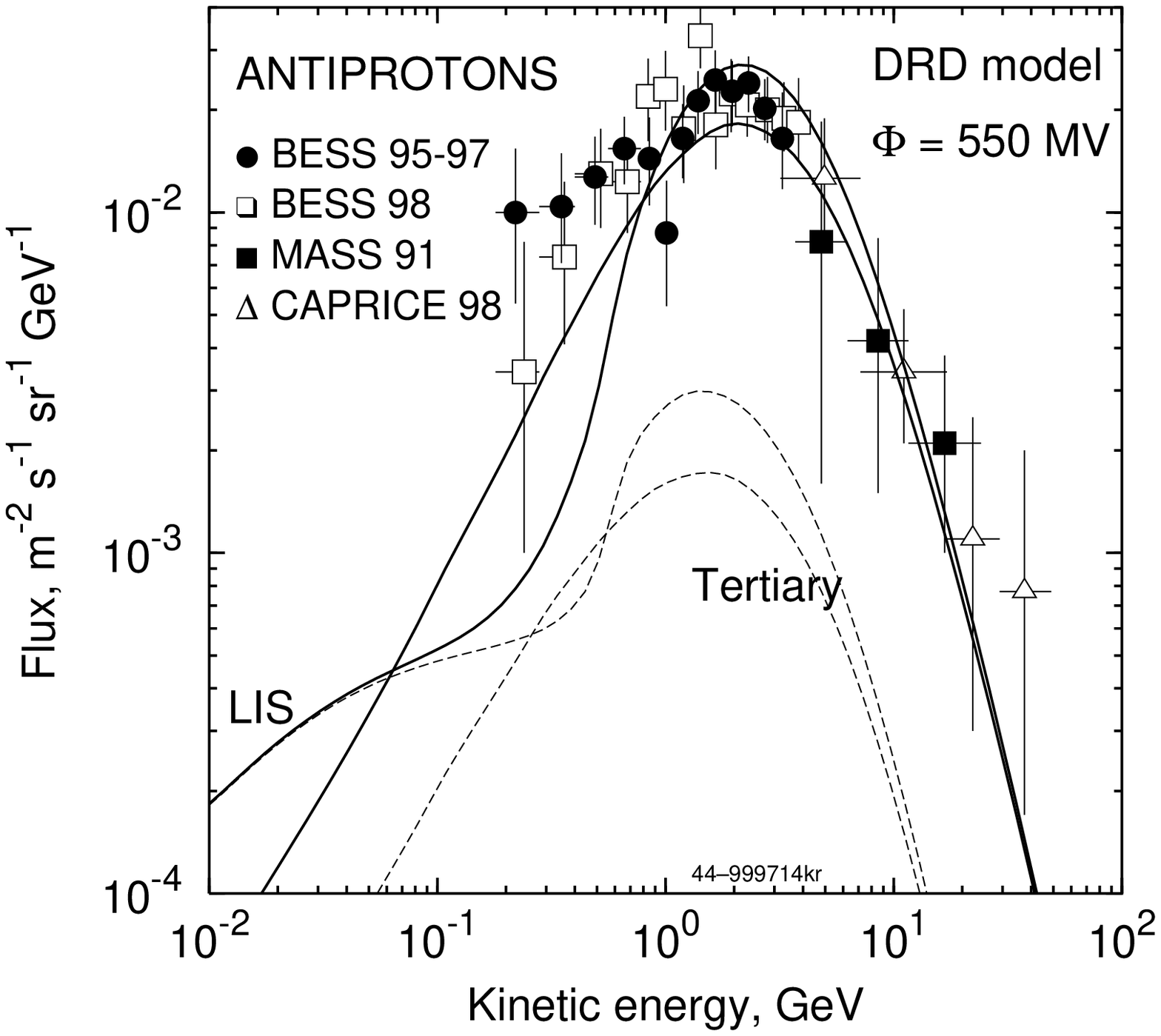}}
\caption{Antiproton flux as calculated in plain diffusion model (PD model),
reacceleration model (RD model), and
diffusive reacceleration with damping model (DRD model). 
Upper curve -- LIS, lower -- modulated to 550 MV.
Data: BESS 95-97 \citep{Orito00}, BESS 98 \citep{Asaoka02},
MASS 91 \citep{basini99}, CAPRICE 98 \citep{boezio01}.\label{pbars}}
\vskip 1\baselineskip
\end{figure}

\begin{figure}
\includegraphics[width=3in]{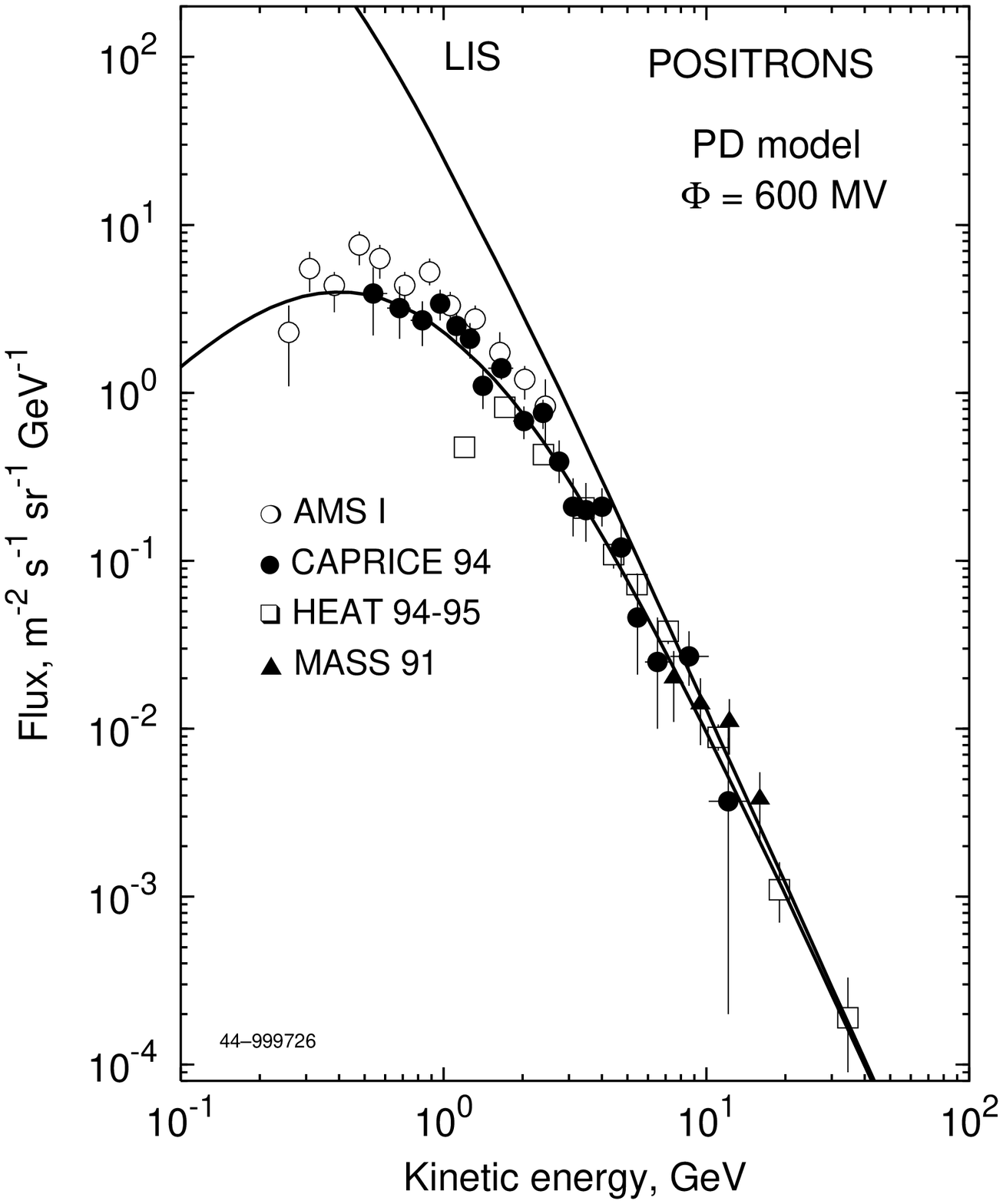}\hfill
\includegraphics[width=3in]{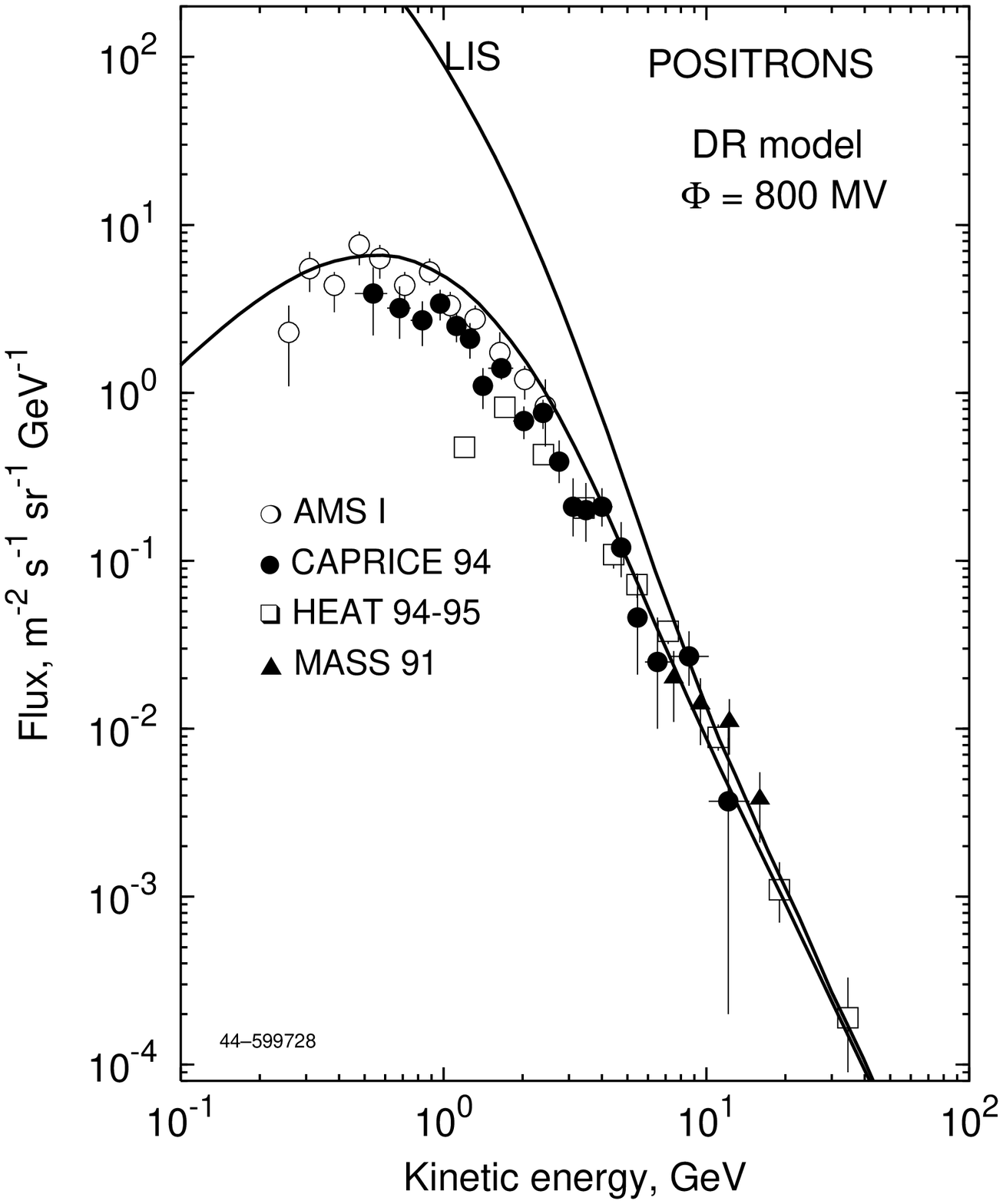}\vfill
\centerline{\includegraphics[width=3in]{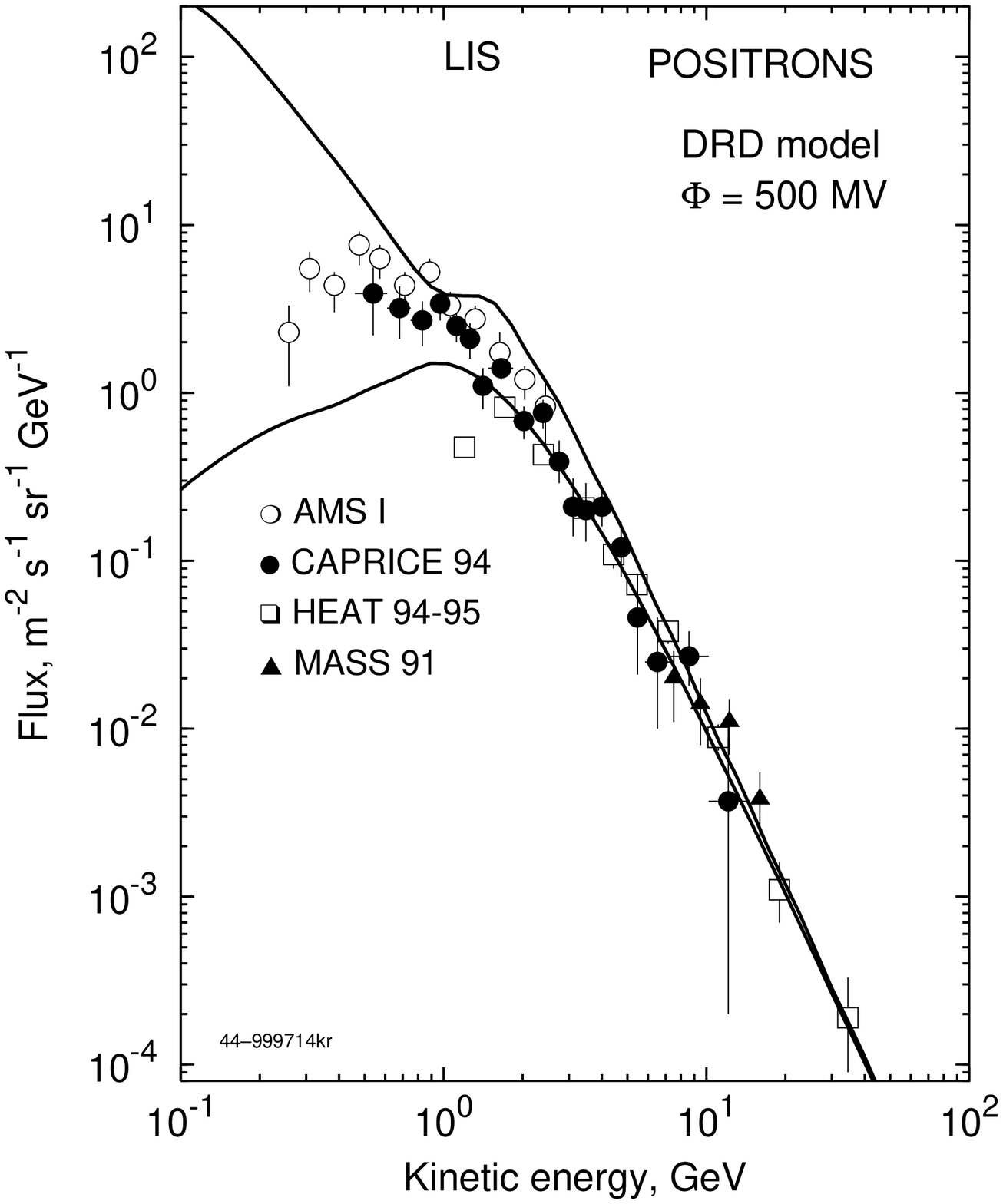}}
\caption{Positron flux as calculated in plain diffusion model (PD model),
reacceleration model (RD model), and
diffusive reacceleration with damping model (DRD model). 
Upper curve -- LIS, lower -- modulated. Data:
AMS-I \citep{leptons_ams}, 
CAPRICE 94 \citep{Boez00}, 
HEAT 94-95 \citep{duvernois01},
MASS 91 \citep{grimani02}.
\label{pos}}
\vskip 1\baselineskip
\end{figure}

\begin{figure}
\includegraphics[width=2.7in]{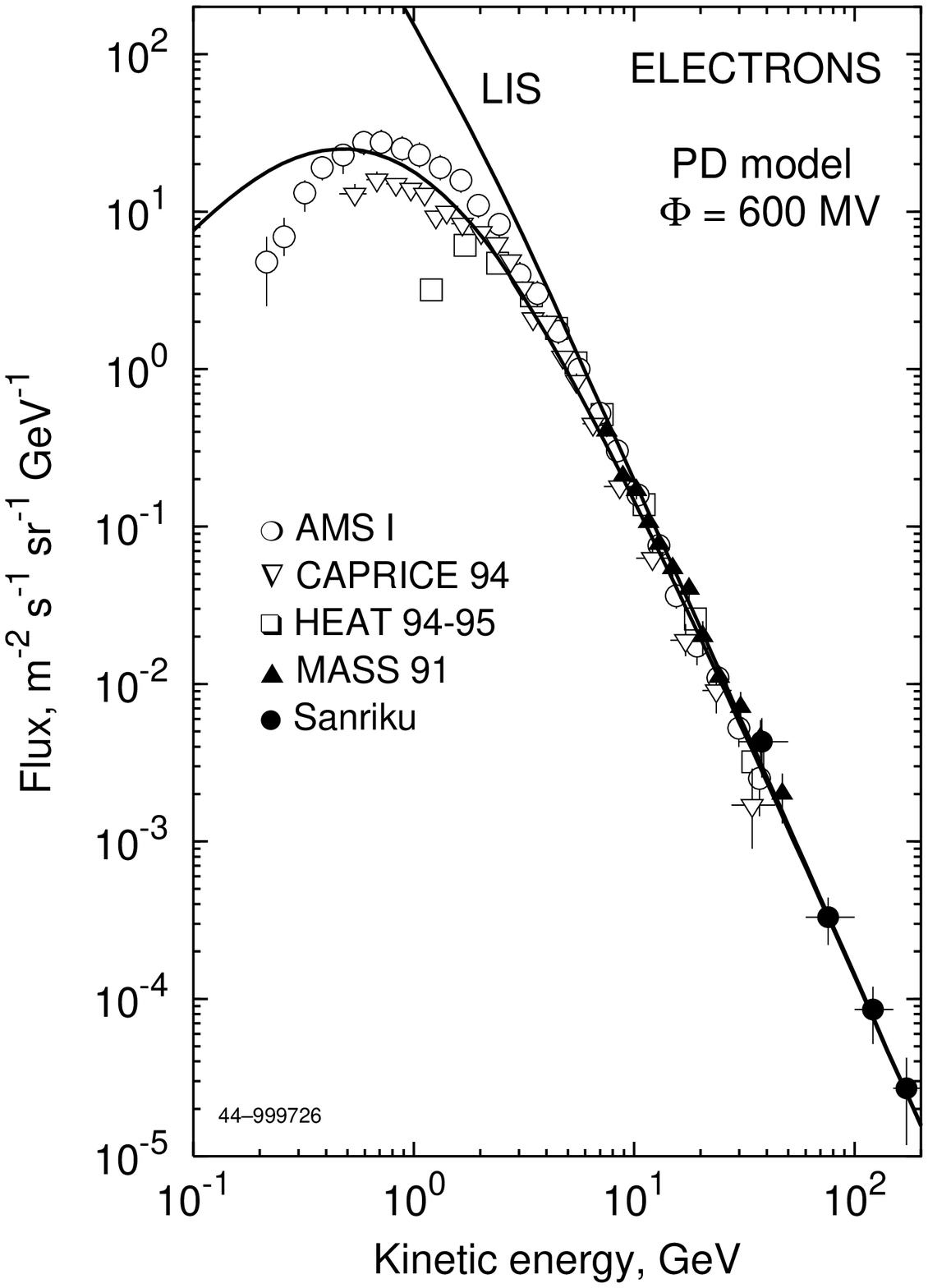}\hfill
\includegraphics[width=2.7in]{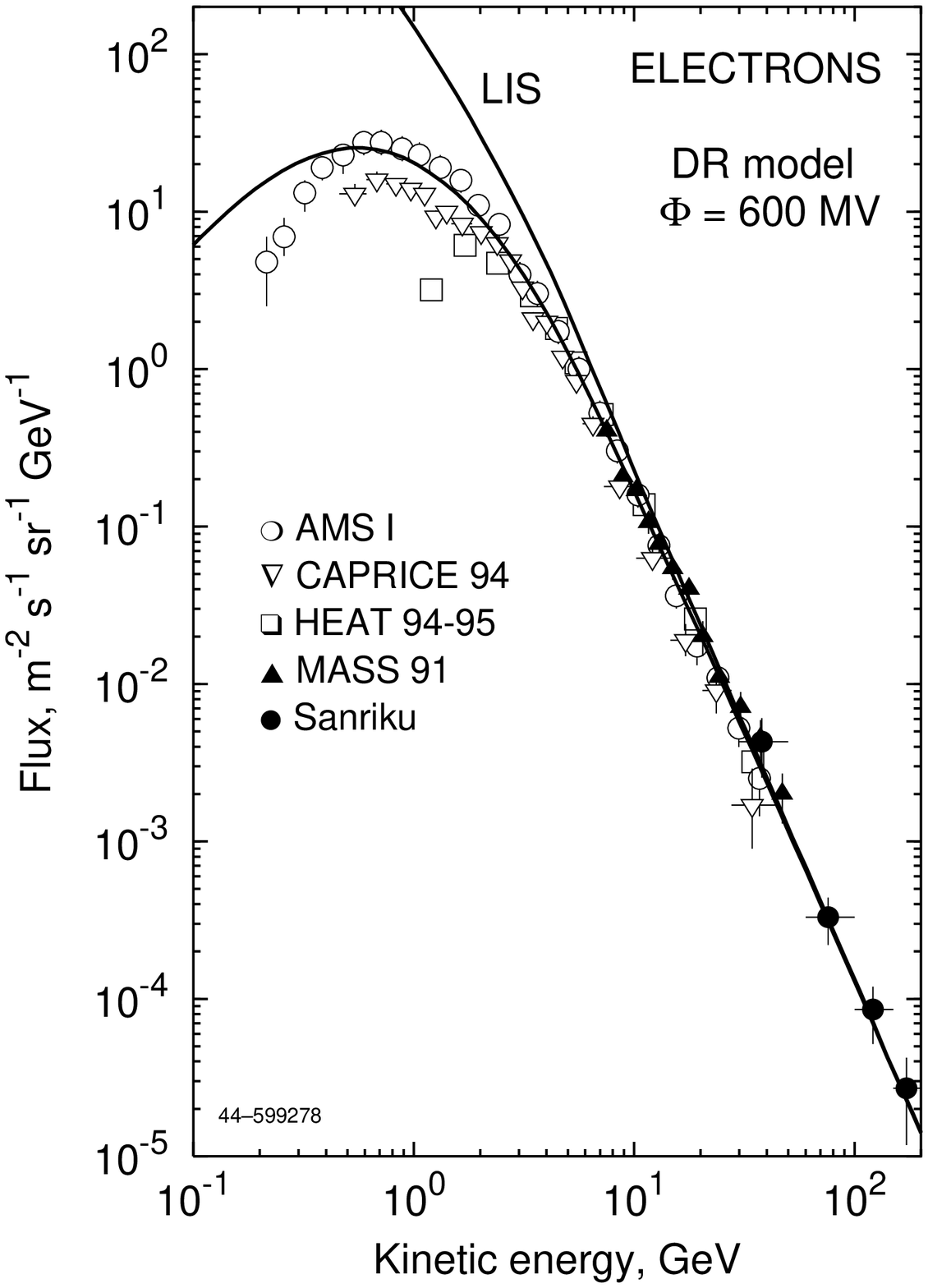}\vfill
\centerline{\includegraphics[width=2.7in]{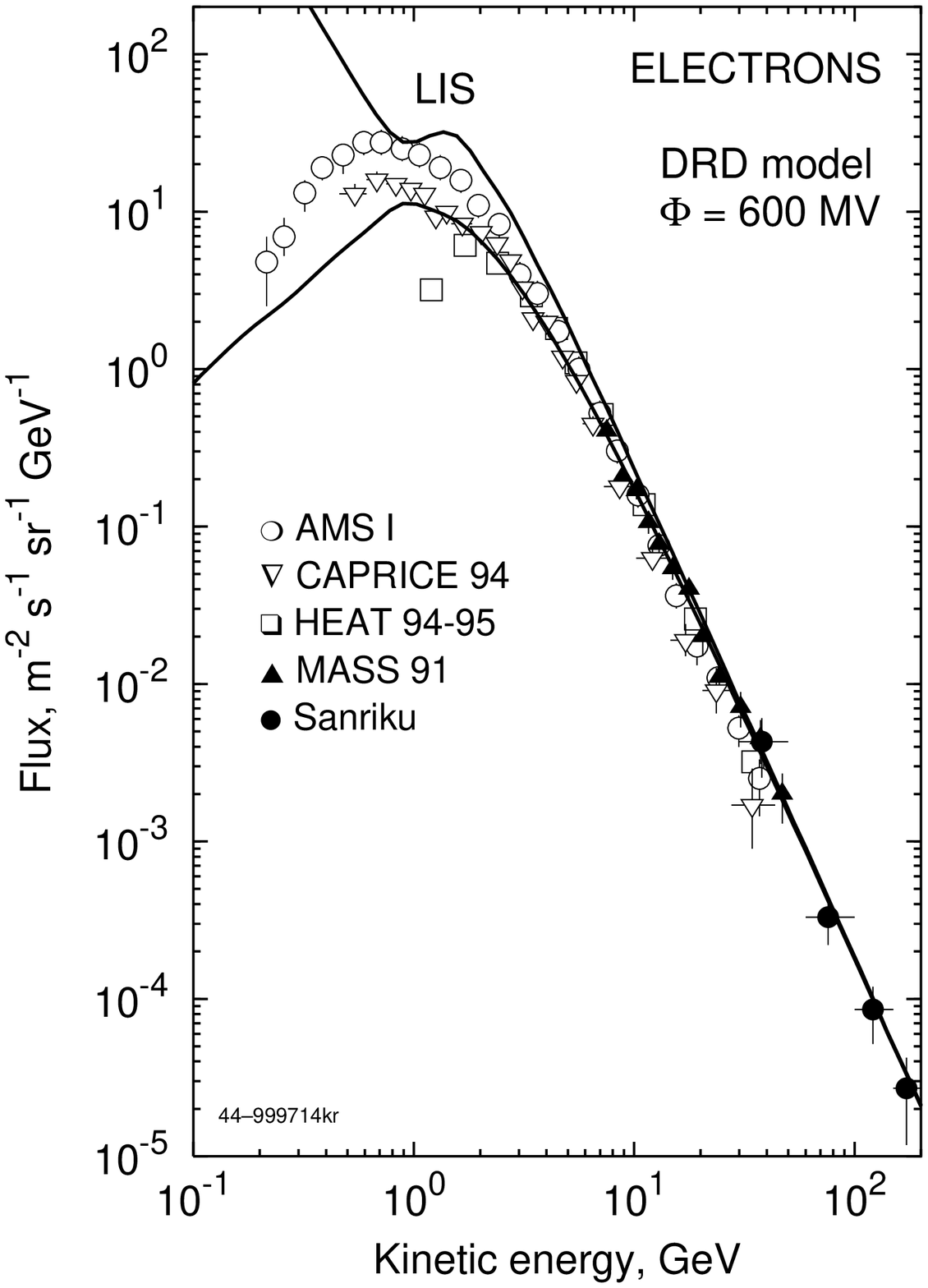}}
\caption{Electron flux (primary plus secondary) as calculated in plain 
diffusion model (PD model), reacceleration model (RD model), and
diffusive reacceleration with damping model (DRD model). 
Upper curve -- LIS, lower -- modulated to 600 MV. Data: 
AMS-I \citep{leptons_ams},
CAPRICE 94 \citep{Boez00}, 
HEAT 94-95 \citep{duvernois01},
MASS 91 \citep{grimani02}, 
Sanriku \citep{kobayashi99}.
\label{ele}}
\vskip 1\baselineskip
\end{figure}

The new self-consistent model  of diffusive reacceleration with damping 
(DRD model) is compared with two reference models which were
earlier found to give good fits to cosmic ray data:
the plain diffusion model (PD model) with an ad-hoc break in the diffusion coefficient and
the model with distributed reacceleration (DR model) and   power-law diffusion with no
breaks \citep{M02}. The parameters of these models found from the fit to the observed
spectra of primary and secondary nuclei are summarized in Table~\ref{table1}.
Note that the parameters of the DRD model given in
Table~\ref{table1} are the ``seed'' values used in the iteration procedure. 
The diffusion coefficient after iteration calculations is shown in Fig.~\ref{Dxx}.

The diffusion coefficients in all three models are presented in Fig.~\ref{Dxx}. 
The DR model diffusion coefficient has a weak energy dependence 
with a single index 1/3 as dictated by the assumed Kolmogorov spectrum of
interstellar turbulence. The PD model requires a break in the diffusion 
coefficient  and an additional factor of $\beta^{-3}$
to be able to match the B/C ratio at low energies.
The DRD model diffusion coefficient lies  between these
at high energies and has a sharp increase at ~1 GV.
Of course, the actual mean free path length can not be infinite. We
put an upper limit of $\sim$15 pc based on the estimate of streaming
instability effect that arises at low rigidities and leads to the
generation of additional turbulence which limits cosmic ray escape
from the Galaxy, see discussion below in Section 6.
The selfconsistent
spectrum of waves calculated in the DRD model is shown in Fig.~\ref{waves}. The
spectrum bends downward at wavenumbers $k>6\times10^{-12}$ cm$^{-1}$ 
due to the wave damping on cosmic rays. 

The B/C ratio and Carbon spectra (primary) after modulation
look almost identical in all three models (Figs.~\ref{bc}, \ref{carbon}). 
The modulated proton spectrum (Fig.~\ref{protons}) matches the data
in all three cases after some tuning, while
the interstellar spectra are different.
The most dramatic difference is the reduction of proton flux  in the DRD model 
at low energies, consistent with the steep increase of the diffusion coefficient.
The antiproton spectrum (Fig.~\ref{pbars}) appears to be very
sensitive to the model assumptions and might help to discriminate between
the models. The DR model produces too few antiprotons, a well known
effect \citep{M02}. The PD and DRD models are both consistent with
antiproton measurements in the heliosphere, but predict very different
spectra in the interstellar medium; the interstellar antiproton flux at low energies
($<$600 MeV) in DRD model is an order of magnitude lower than in two other models. 
In both reacceleration models (DR and DRD), the majority of low-energy 
antiprotons come from inelastic scattering (so-called ``tertiary'' antiprotons).

Figs.~\ref{pos} and \ref{ele} show secondary positrons
and primary plus secondary electrons as calculated in all three models.
The spectra are similar in the PD and DR models, while DRD spectra
exhibit lower intensities at low energies.
This may be an observable effect since the models predict different
synchrotron emission spectra (electrons).

\section{Discussion}

 Damping on cosmic rays may terminate the slow Kraichnan-type cascade in
the interstellar medium at $k\sim10^{-12}$ cm$^{-1}$. Our estimates were made
for the level of MHD turbulence which produces the empirical value of cosmic-ray
diffusion coefficient. This finding suggests a possible explanation for the peaks in
secondary/primary nuclei ratios at about 1 GeV/n observed in cosmic rays:
the amplitude of short waves is small because of damping and thus the low
energy particles rapidly exit the Galaxy without producing many secondaries.
There is no other obvious reasons for a sharp cut off in the wave spectrum. If
the concept of MHD turbulence by \citet{Gol95} works for
interstellar turbulence, the MHD waves we are dealing with in this
context are the fast magnetosonic waves. The Alfven waves propagate
predominantly perpendicular to the magnetic field and because of this they do
not significantly scatter cosmic rays. It also explains why radio scintillation
observations show no sign of the termination of electron density fluctuations at
wave numbers between $10^{-14}$ to $10^{-8}$ cm$^{-1}$. According to
\citet{Lit01} these fluctuations are produced by the slow magnetosonic
waves with $k_{\perp}\gg k_{\parallel}$ which are almost not damped on cosmic
rays. An alternative explanation is that the wave damping on cosmic rays and
the radio scintillations mainly occur in separate regions of the
interstellar medium (see below in the discussion on the ``sandwich'' model of
cosmic ray propagation in the Galaxy). Some minor contribution to the observed
scintillations is possible from fast
magnetosonic waves interacting with energetic particles, 
and in this respect it is of interest that the
observations may need an enhancement in the power on large ``refractive''
scales $10^{13}-10^{14}$ cm relative to the power on small ``diffractive''
scales $10^{9}-10^{10}$ cm \citep{Lam00}. This may indicate the
cutoff of the spectrum of fast magnetosonic waves due to cosmic ray action.

While the mere fact of a wave spectrum steepening under the action of damping
on cosmic rays can be described by a simple eq.~(\ref{eq9}) with some
characteristic time for nonlinear wave interactions $T_{nl}[W(k),k]$, the exact
form of the function $W(k)$ at large $k$ where the damping is significant depends
critically  on the form of the equation for waves. It involves in
particular to the vanishing of $W(k)$ at some $k_{\ast}$ ($\sim$$10^{-12}$
cm$^{-1}$) and the corresponding singularity of $D(p)$ at some $p_{\ast}$
($\sim$1 GV) found in our calculations. Less significant in this sense is our
approximation of the resonant wave number in eq.~(\ref{eq5}), 
which does not include the particle pitch angle  in 
an explicit form. (If it were included, the term with
$g$ in eq.~[\ref{eq9}] would have a third integration over the pitch angle.) We note
however that this effect was included in the derivation of eq.~[\ref{eq10}] for the
attenuation rate.

In the context of the approximations adopted  in the present work, the problem of
cosmic-ray transport at $p<p_{\ast}$ arises. The free streaming of cosmic rays
from the Galaxy leads to an instability and to the growth of waves which scatter
particles and thus slow down the streaming, see e.g. \citet{Ber90}. 
We shall consider the processes at low energies in a separate work.
Here we note only that the streaming instability develops above the Galactic
disk at $D(1$\textrm{\ GV}$)\gtrsim V_{\mathrm{a}}H\sim3\times10^{29}$
cm$^{2}$/s (if the wave damping is absent) and leads to the
diffusive-convective transport of cosmic rays. The given estimate of the
diffusion coefficient at 1 GV follows from the condition of cosmic ray streaming
instability $U_{cr}>V_{a}$ where $U_{cr}$ is the bulk velocity of cosmic ray gas.
The bulk velocity is $U_{cr}\approx\delta_{cr}c$ where the cosmic ray anisotropy
perpendicular to galactic disk is $\delta_{cr}\approx D/cH$. The Alfven velocity
in galactic halo is about $2\times10^{7}$ cm/s. It is important that the
magnetic rigidity $1$ GV corresponds to a   kinetic energy $0.43$ GeV for
protons, and $0.13$ GeV/n for nuclei with charge to mass ratio $Z/A=1/2$. The
Galactic spectrum of cosmic-ray protons and nuclei at such low energies can not be derived from
direct observations at the Earth because of strong modulation in the solar wind.

 Dissipation  other than on  cosmic rays has been  neglected in the present work,
though it may completely destroy the MHD cascade or considerably change its
angular distribution and thus affect $D(p)$ in a large part of the
interstellar medium, see \citet{McI77}, \citet{Ces80},
\citet{Yan04}. The region of the cosmic ray halo is the most ``safe'' in this sense
\citep{Yan04}. In particular,     dissipation on ion-neutral collisions
may destroy MHD turbulence in the Galactic disk but not in the halo where
neutrals are absent. We  then  come to the ``sandwich'' model for cosmic-ray
 propagation, with different diffusion coefficients in the Galactic disk,
$D_{\mathrm{g}}$, and in the halo, $D_{\mathrm{h}}$ (it is assumed here that
some scattering is present in the Galactic disk and the diffusion
approximation works). In this case, the mean matter thickness traversed by
cosmic rays does not depend on diffusion inside the disk even at zero gas
density in the halo: $X=\frac{\mu_{\mathrm{g}}v}{2}\left(  \frac
{h}{2D_{\mathrm{g}}}+\frac{H}{D_{\mathrm{h}}}\right)  \approx\mu_{\mathrm{g}%
}vH/(2D_{\mathrm{h}})\mathrm{\ }$ \citep{Gin76}. The energy
dependence of secondary to primary ratios in cosmic rays is determined by the
diffusion coefficient in the cosmic ray halo where the model developed in the
present paper is applied.

The estimate based on the empirical value of the diffusion coefficient for GeV
particles (see Section 1) gives the level of turbulence at the principal scale
$\delta B_{\mathrm{tot}}^{2}/B^{2}\sim$\ $0.03$ for a Kraichnan-type spectrum
$W(k)\varpropto k^{-3/2}$, and $\delta B_{\mathrm{tot}}^{2}/B^{2}\sim1$ for a
Kolmogorov-type spectrum $W(k)\varpropto k^{-5/3}$, if $k_{\mathrm{L}}%
=10^{-21}$ cm$^{-1}$. At the same time, the data on Faraday rotation angles
favor the Kolmogorov spectrum with $\delta B_{\mathrm{tot}}^{2}/B^{2}\sim1$
and $k_{\mathrm{L}}=10^{-21}$ cm$^{-1}$. The cascades of Alfven waves (with the
scaling $k^{-5/3}$) and the fast magnetosonic waves ($k^{-3/2}$) are
independent in the \citet{Gol95} model of MHD turbulence, and
the amplitude of  Alfven wave cascade may dominate at the principle scale.
Also, in the ``sandwich'' model described in a previous paragraph, the
turbulence, which determines the confinement of cosmic rays in the Galaxy, is
distributed in the halo of size $H\sim4$ kpc whereas the observations of
interstellar turbulence refer to the Galactic disk and the adjacent region,
where it can be much stronger.

\section{Conclusions}

On the whole, the empirical diffusion model for cosmic rays with energies from
$10^{8}$ to $10^{17}$ eV implies the presence of random magnetic field with an
extended power law spectrum of fluctuations $W(k)\varpropto k^{-2+a}$,
$a\lesssim0.5$\ at wave numbers from $3\times10^{-12}$ to $10^{-20}$ cm$^{-1}$.
The existence of such a turbulence spectrum in the interstellar
medium seems confirmed by various astronomical observations. It should be
emphasized that this does not prove the existence of a spectral cascade or
spectral transfer throughout this enormous wavenumber range, however tempting
that conclusion may be. Among other puzzles, an oddity is the absence of a
spectral feature on spatial scales where ion-neutral collisional processes
should be most pronounced. The two special cases of turbulence spectrum with
$a=1/3$ and $a=1/2$ which correspond to the Kolmogorov and the Kraichnan
spectra respectively are used in popular versions of the diffusion model as
described in Section 5. In both cases, as the present work has demonstrated,
one can get a satisfactory fit to the data on energy spectra of secondary and
primary nuclei if   account is taken for wave damping on cosmic rays for a
slow Kraichnan cascade. The fit can be obtained either in the DR model with
reacceleration on the Kolmogorov spectrum with no significant effect of
cosmic ray damping (because the Kolmogorov cascade is fast) or in the DRD model
with relatively weak reacceleration on the Kraichnan spectrum which is
significantly modified by cosmic ray damping. Some problems still remain
to be solved. The main
difficulties with the Kolmogorov spectrum are, firstly, the contradiction with the
leading theory of MHD turbulence where this spectrum is associated with 
perpendicular propagating Alfven waves which almost do not scatter cosmic-ray
particles; and secondly, the low flux of antiprotons characteristic of             
models with relatively strong reacceleration. A major problem of concern for
diffusion on a Kraichnan spectrum is the relatively strong dependence of
diffusion on energy which leads to an unacceptably large anisotropy of cosmic
rays especially above $10^{14}$ eV, see \citet{Jon01}, \citet{Ptu03b}.

Let us emphasize again that the models of cosmic ray propagation discussed in
the present paper assume that the MHD turbulence required for cosmic-ray
scattering is produced by some external sources. An alternative is the
Galactic wind model by \citet{Zir96} and \citet{Ptu97}.
In this model, the cosmic-ray pressure drives a wind with a frozen-in regular
magnetic field which is shaped into huge spirals (the radius of the Galactic
wind cavity is about $300$ Kpc). The MHD turbulence in the wind is created by
the streaming instability of cosmic rays moving predominantly along the
regular magnetic field lines outward from the Galactic disk. The level of turbulence
is regulated by the nonlinear Landau damping on thermal ions. The model
explains well the cosmic-ray data up to ultra-high energies $\sim$$10^{17}$ eV
with the exception of the observed low anisotropy (about the same difficulty with
anisotropy as occurs in the diffusion model with a given Kraichnan spectrum).
As regards interpretation of cosmic-ray observations, a preference cannot yet
be given to any of the models of cosmic-ray transport in the Galaxy discussed above. 

\acknowledgments

V.\ S.\ P.\ and F.\ C.\ J.\ acknowledge
partial support from NASA Astrophysics Theory Program (ATP) grants. V.\ S.\ P.\ and
V.\ N.\ Z.\ were supported by the RFBR grant at IZMIRAN. I.\ V.\ M.\  acknowledges
partial support from NASA Astrophysics Theory Program (ATP) grant and NASA
Astronomy and Physics Research and Analysis Program (APRA) grant.

\end{document}